\documentclass[%
aps, prapplied,
amsmath,amssymb,
onecolumn,
preprint,
]{revtex4-2}

\usepackage[utf8]{inputenc} % Caracteres con acentos.
\ifdefined\HelvFont
\usepackage{helvet}
\fi
\usepackage{geometry}			%%% Control de los márgenes
\usepackage{hyperref} 		%%%
\usepackage[nomain,acronym,toc]{glossaries}
\usepackage{graphicx}
\usepackage{tabularx}
\usepackage{color}				%%% Colores
\usepackage[dvipsnames]{xcolor}				%%% Colores
\usepackage{amsfonts}
\usepackage{amsthm}
\usepackage{fontenc} %% %%
\usepackage{textcomp} %% %%
\usepackage{braket} %% %%
\usepackage{wasysym}
\usepackage{mathtools}
\usepackage{dcolumn} %% %%
\usepackage{lmodern} 			%%% Type1-font for non-english texts and characters
\usepackage{adjustbox}
\usepackage{multirow}
\usepackage{pgfplots}			%%% 
\usepackage{tikz}					%%%
\usetikzlibrary{shapes}
\usetikzlibrary{calc}
\usetikzlibrary{positioning}
\usepgfplotslibrary{fillbetween}
\usepgfplotslibrary{colormaps} 
\usetikzlibrary{shapes.geometric}
\usetikzlibrary{arrows}
\usepackage{circuitikz}
\usepackage{tikzscale}		%%%
\usepackage{latexsym} 		%%% Símbolos
\usepackage{float}				%%%
\usepackage{fancyhdr} 		%%Fancy headings
\usepackage{titlesec} 		%% http://www.ctex.org/documents/packages/layout/titlesec.pdf
\usepackage{listings}
\usepackage{marginnote}
\usepackage[normalem]{ulem}
\usepackage{etoolbox}
\usepackage{appendix}
\usepackage{makecell}
\usepackage{relsize}

\usepackage[eulergreek]{sansmath}

\ifdefined\ExtComp
\usepgfplotslibrary{external}
\fi

\ifdefined\HelvFont

\usepackage{mathastext}
\pgfplotsset{every tick label/.append style={font={\sffamily}}}
\pgfplotsset{every node/.style={/utils/exec={\sfdefault}}}
\fi

\definecolor{naranja_1}{rgb}{0.70,0.25,0}
\definecolor{gris-azul}{rgb}{0.27,0.33,0.415}  
\definecolor{azul-gris}{rgb}{0.192,0.278,0.4}  
\definecolor{verde-chulo}{rgb}{0.25,0.5,0.5}  
\definecolor{verde-chulo2}{rgb}{0,0.5,0.4} 

\definecolor{color1}{rgb}{1, 0, 0}%
\definecolor{color2}{rgb}{0, 0, 1}%
\definecolor{color3}{rgb}{0, 1, 0}%
\definecolor{color4}{rgb}{1, 0, 1}%
\definecolor{color5}{rgb}{0, 1, 1}%

\definecolor{mycolor11}{RGB}{166,189,219}%
\definecolor{mycolor6}{RGB}{116,169,207}%
\definecolor{mycolor1}{RGB}{54,144,192}%
\definecolor{mycolor16}{RGB}{5,112,176}%
\definecolor{mycolor21}{RGB}{4,90,141}%
\definecolor{mycolor17}{RGB}{252, 146, 114}%
\definecolor{mycolor7}{RGB}{251, 106, 74}%
\definecolor{mycolor12}{RGB}{239, 59, 44}%
\definecolor{mycolor2}{RGB}{203, 24, 29}%
\definecolor{mycolor22}{RGB}{165, 15, 21}%
\definecolor{mycolor18}{RGB}{250, 159, 181}%
\definecolor{mycolor8}{RGB}{247, 104, 161}%
\definecolor{mycolor13}{RGB}{221, 52, 151}%
\definecolor{mycolor5}{RGB}{174, 1, 126}%
\definecolor{mycolor23}{RGB}{122, 1, 119}%
\definecolor{mycolor19}{RGB}{254,196,79}%
\definecolor{mycolor9}{RGB}{254,153,41}%
\definecolor{mycolor14}{RGB}{236,112,20}%
\definecolor{mycolor4}{RGB}{204,76,2}%
\definecolor{mycolor24}{RGB}{153,52,4}%
\definecolor{mycolor20}{RGB}{153, 216, 201}%
\definecolor{mycolor10}{RGB}{102, 194, 164}%
\definecolor{mycolor15}{RGB}{65, 174, 118}%
\definecolor{mycolor3}{RGB}{35, 139, 69}%
\definecolor{mycolor25}{RGB}{0, 109, 44}%

\pgfplotscreateplotcyclelist{molecColorList}{
color1, line width=0.8pt\\
color2, line width=0.8pt\\
color3, line width=0.8pt\\
color4, line width=0.8pt\\
color5, line width=0.8pt\\
color1, mark=o\\
color2, mark=o\\
color3, mark=o\\
color4, mark=o\\
color5, mark=o\\
}

\pgfplotsset{%
	colormap={parula}{
	rgb255=(53,42,135)
	rgb255=(15,92,221)
	rgb255=(18,125,216)
	rgb255=(7,156,207)
	rgb255=(21,177,180)
	rgb255=(89,189,140)
	rgb255=(165,190,107)
	rgb255=(225,185,82)
	rgb255=(252,206,46)
	rgb255=(249,251,14)
	},
	colormap/bluered,
	colormap/jet,
	colormap={RdBu}{
	rgb255=(103,1,31)
	rgb255=(176,23,42)
	rgb255=(213,95,26)
	rgb255=(243,163,128)
	rgb255=(253,218,198)
	rgb255=(255,255,255)
	rgb255=(209,229,240)
	rgb255=(144,196,221)
	rgb255=(66,146,194)		
	rgb255=(32,100,170)		
	rgb255=(5,48,167)
	},
	colormap={linBR}{
	rgb255=(0,0,255)
	rgb255=(27,0,228)
	rgb255=(54,0,201)
	rgb255=(80,0,175)
	rgb255=(107,0,148)
	rgb255=(134,0,121)
	rgb255=(161,0,94)
	rgb255=(188,0,67)
	rgb255=(215,0,40)
	rgb255=(241,0,14)
	rgb255=(255,0,0)
	},
	colormap={linBR2}{
	rgb255=(0,0,255)
	rgb255=(54,0,201)
	rgb255=(107,0,148)
	rgb255=(161,0,94)
	rgb255=(215,0,40)
	rgb255=(255,0,0)
	},
	colormap={linBR3}{
	rgb255=(0,0,255)
	rgb255=(6,0,249)
	rgb255=(12,0,243)
	rgb255=(18,0,237)
	rgb255=(24,0,231)
	rgb255=(30,0,225)
	rgb255=(36,0,219)
	rgb255=(42,0,213)
	rgb255=(48,0,207)
	rgb255=(54,0,201)
	rgb255=(60,0,195)
	rgb255=(66,0,189)
	rgb255=(72,0,183)
	rgb255=(78,0,177)
	rgb255=(84,0,171)
	rgb255=(90,0,165)
	rgb255=(96,0,159)
	rgb255=(102,0,153)
	rgb255=(108,0,147)
	rgb255=(114,0,141)
	rgb255=(120,0,135)
	rgb255=(126,0,129)
	rgb255=(132,0,123)
	rgb255=(138,0,117)
	rgb255=(144,0,111)
	rgb255=(150,0,105)
	rgb255=(156,0,99)
	rgb255=(162,0,93)
	rgb255=(168,0,87)
	rgb255=(174,0,81)
	rgb255=(180,0,75)
	rgb255=(186,0,69)
	rgb255=(192,0,63)
	rgb255=(198,0,57)
	rgb255=(204,0,51)
	rgb255=(210,0,45)
	rgb255=(216,0,39)
	rgb255=(222,0,33)
	rgb255=(228,0,27)
	rgb255=(234,0,21)
	rgb255=(240,0,15)
	rgb255=(246,0,9)
	rgb255=(255,0,0)
	},
	colormap={linBRmain}{
		rgb255=(0,0,255)
		rgb255=(14,0,241)%
		rgb255=(27,0,228)
		rgb255=(40,0,215)%
		rgb255=(54,0,201)
		rgb255=(67,0,188)%
		rgb255=(80,0,175)
		rgb255=(94,0,161)%
		rgb255=(107,0,148)
		rgb255=(121,0,134)%
		rgb255=(134,0,121)%
		rgb255=(148,0,107)
		rgb255=(161,0,94)%
		rgb255=(175,0,80)
		rgb255=(188,0,67)%
		rgb255=(201,0,54)
		rgb255=(215,0,40)%
		rgb255=(228,0,27)
		rgb255=(241,0,14)%
		rgb255=(255,0,0)
	},
	colormap={cmap1}{
		rgb255=(53,42,135)
		rgb255=(55,47,148)
		rgb255=(55,53,160)
		rgb255=(55,58,171)
		rgb255=(53,62,180)
		rgb255=(51,67,189)
		rgb255=(47,71,196)
		rgb255=(43,76,203)
		rgb255=(38,80,208)
		rgb255=(32,84,213)
		rgb255=(25,88,217)
		rgb255=(16,92,221)
		rgb255=(5,95,223)
		rgb255=(0,99,224)
		rgb255=(0,102,225)
		rgb255=(0,105,224)
		rgb255=(0,108,223)
		rgb255=(0,111,222)
		rgb255=(4,113,221)
		rgb255=(9,116,219)
		rgb255=(14,119,218)
		rgb255=(17,122,217)
		rgb255=(18,124,216)
		rgb255=(18,127,216)
		rgb255=(17,130,215)
		rgb255=(16,133,215)
		rgb255=(15,136,214)
		rgb255=(14,139,214)
		rgb255=(12,142,213)
		rgb255=(11,145,212)
		rgb255=(9,148,212)
		rgb255=(8,150,210)
		rgb255=(7,153,209)
		rgb255=(7,155,208)
		rgb255=(7,158,206)
		rgb255=(5,160,204)
		rgb255=(2,162,202)
		rgb255=(0,164,200)
		rgb255=(0,166,197)
		rgb255=(0,168,195)
		rgb255=(0,170,192)
		rgb255=(0,172,190)
		rgb255=(0,173,187)
		rgb255=(7,175,184)
		rgb255=(17,176,181)
		rgb255=(26,178,178)
		rgb255=(33,179,175)
		rgb255=(39,181,172)
		rgb255=(44,182,168)
		rgb255=(50,183,164)
		rgb255=(56,184,160)
		rgb255=(56,184,160)
		rgb255=(85,189,142)
		rgb255=(121,190,126)
		rgb255=(155,190,112)
		rgb255=(185,188,99)
		rgb255=(213,185,88)
		rgb255=(233,187,76)
		rgb255=(245,196,60)
		rgb255=(253,209,42)
		rgb255=(255,228,24)
		rgb255=(249,251,14)
	},
	colormap={linBRmain100}{
		rgb255=(0,0,255)
		rgb255=(2,0,253)
		rgb255=(4,0,251)
		rgb255=(6,0,249)
		rgb255=(8,0,247)
		rgb255=(10,0,245)
		rgb255=(12,0,243)
		rgb255=(14,0,241)
		rgb255=(16,0,239)
		rgb255=(18,0,237)
		rgb255=(20,0,235)
		rgb255=(22,0,233)
		rgb255=(24,0,231)
		rgb255=(26,0,229)
		rgb255=(28,0,227)
		rgb255=(30,0,225)
		rgb255=(32,0,223)
		rgb255=(34,0,221)
		rgb255=(36,0,219)
		rgb255=(38,0,217)
		rgb255=(40,0,215)
		rgb255=(42,0,213)
		rgb255=(44,0,211)
		rgb255=(46,0,209)
		rgb255=(48,0,207)
		rgb255=(50,0,205)
		rgb255=(52,0,203)
		rgb255=(54,0,201)
		rgb255=(56,0,199)
		rgb255=(58,0,197)
		rgb255=(60,0,195)
		rgb255=(62,0,193)
		rgb255=(64,0,191)
		rgb255=(66,0,189)
		rgb255=(68,0,187)
		rgb255=(70,0,185)
		rgb255=(72,0,183)
		rgb255=(74,0,181)
		rgb255=(76,0,179)
		rgb255=(78,0,177)
		rgb255=(80,0,175)
		rgb255=(82,0,173)
		rgb255=(84,0,171)
		rgb255=(86,0,169)
		rgb255=(88,0,167)
		rgb255=(90,0,165)
		rgb255=(92,0,163)
		rgb255=(94,0,161)
		rgb255=(96,0,159)
		rgb255=(98,0,157)
		rgb255=(100,0,155)
		rgb255=(102,0,153)
		rgb255=(104,0,151)
		rgb255=(106,0,149)
		rgb255=(108,0,147)
		rgb255=(110,0,145)
		rgb255=(112,0,143)
		rgb255=(114,0,141)
		rgb255=(116,0,139)
		rgb255=(118,0,137)
		rgb255=(120,0,135)
		rgb255=(122,0,133)
		rgb255=(124,0,131)
		rgb255=(126,0,129)
		rgb255=(128,0,127)
		rgb255=(130,0,125)
		rgb255=(132,0,123)
		rgb255=(134,0,121)
		rgb255=(136,0,119)
		rgb255=(138,0,117)
		rgb255=(140,0,115)
		rgb255=(142,0,113)
		rgb255=(144,0,111)
		rgb255=(146,0,109)
		rgb255=(148,0,107)
		rgb255=(150,0,105)
		rgb255=(152,0,103)
		rgb255=(154,0,101)
		rgb255=(156,0,99)
		rgb255=(158,0,97)
		rgb255=(160,0,95)
		rgb255=(162,0,93)
		rgb255=(164,0,91)
		rgb255=(166,0,89)
		rgb255=(168,0,87)
		rgb255=(170,0,85)
		rgb255=(172,0,83)
		rgb255=(174,0,81)
		rgb255=(176,0,79)
		rgb255=(178,0,77)
		rgb255=(180,0,75)
		rgb255=(182,0,73)
		rgb255=(184,0,71)
		rgb255=(186,0,69)
		rgb255=(188,0,67)
		rgb255=(190,0,65)
		rgb255=(192,0,63)
		rgb255=(194,0,61)
		rgb255=(196,0,59)
		rgb255=(198,0,57)
		rgb255=(200,0,55)
		rgb255=(202,0,53)
		rgb255=(204,0,51)
		rgb255=(206,0,49)
		rgb255=(208,0,47)
		rgb255=(210,0,45)
		rgb255=(212,0,43)
		rgb255=(214,0,41)
		rgb255=(216,0,39)
		rgb255=(218,0,37)
		rgb255=(220,0,35)
		rgb255=(222,0,33)
		rgb255=(224,0,31)
		rgb255=(226,0,29)
		rgb255=(228,0,27)
		rgb255=(230,0,25)
		rgb255=(232,0,23)
		rgb255=(234,0,21)
		rgb255=(236,0,19)
		rgb255=(238,0,17)
		rgb255=(240,0,15)
		rgb255=(242,0,13)
		rgb255=(244,0,11)
		rgb255=(246,0,9)
		rgb255=(248,0,7)
		rgb255=(250,0,5)
		rgb255=(252,0,3)
		rgb255=(254,0,1)
	},
	colormap={linBR9}{
		rgb255=(0,0,255)
		rgb255=(28,0,227)
		rgb255=(56,0,199)
		rgb255=(84,0,171)
		rgb255=(112,0,143)
		rgb255=(140,0,115)
		rgb255=(168,0,87)
		rgb255=(196,0,59)
		rgb255=(224,0,31)
		rgb255=(252,0,3)
	},
	colormap={linBR5}{
		rgb255=(0,0,255)
		rgb255=(51,0,204)
		rgb255=(102,0,153)
		rgb255=(153,0,102)
		rgb255=(204,0,51)
	},
	colormap={linBR21}{	
	rgb255=(0,0,255)
	rgb255=(12,0,243)
	rgb255=(24,0,231)
	rgb255=(36,0,219)
	rgb255=(48,0,207)
	rgb255=(60,0,195)
	rgb255=(72,0,183)
	rgb255=(84,0,171)
	rgb255=(96,0,159)
	rgb255=(108,0,147)
	rgb255=(120,0,135)
	rgb255=(132,0,123)
	rgb255=(144,0,111)
	rgb255=(156,0,99)
	rgb255=(168,0,87)
	rgb255=(180,0,75)
	rgb255=(192,0,63)
	rgb255=(204,0,51)
	rgb255=(216,0,39)
	rgb255=(228,0,27)
	rgb255=(240,0,15)
	rgb255=(252,0,3)
	},
	colormap={newBWR12}{
	rgb255=(0,0,255)
	rgb255=(51,51,255)
	rgb255=(102,102,255)
	rgb255=(153,153,255)
	rgb255=(204,204,255)
	rgb255=(255,255,255)
	rgb255=(255,204,204)
	rgb255=(255,153,153)
	rgb255=(255,102,102)
	rgb255=(255,51,51)
	rgb255=(255,0,0)
	},
	colormap={BlueScale3}{
	rgb255=(34, 108, 166)
	rgb255=(72, 145, 200)
	rgb255=(132,186, 224)
	rgb255=(0,39,82)
	},
	colormap={natMap30}{
		rgb255=(0,39,82)
		rgb255=(13,65,113)
		rgb255=(26,91,145)
		rgb255=(43,125,186)
		rgb255=(64,139,176)
		rgb255=(92,157,164)
		rgb255=(113,170,154)
		rgb255=(137,186,147)
		rgb255=(156,197,142)
		rgb255=(177,210,142)
		rgb255=(189,215,154)
		rgb255=(201,221,165)
		rgb255=(215,227,178)
		rgb255=(226,231,187)
		rgb255=(239,237,197)
		rgb255=(245,232,192)
		rgb255=(245,215,168)
		rgb255=(245,202,149)
		rgb255=(245,186,125)
		rgb255=(245,173,107)
		rgb255=(245,161,88)
		rgb255=(238,141,76)
		rgb255=(231,126,71)
		rgb255=(222,105,64)
		rgb255=(219,91,68)
		rgb255=(215,74,74)
		rgb255=(212,61,79)
		rgb255=(174,48,58)
		rgb255=(146,38,41)
		rgb255=(118,28,25)
	},
	colormap={natBlueRed30}{
		rgb255=(0,39,82)
		rgb255=(13,51,91)
		rgb255=(26,63,101)
		rgb255=(44,79,114)
		rgb255=(57,91,123)
		rgb255=(75,107,136)
		rgb255=(88,119,146)
		rgb255=(106,135,159)
		rgb255=(120,147,168)
		rgb255=(137,163,181)
		rgb255=(151,175,191)
		rgb255=(164,187,200)
		rgb255=(182,203,213)
		rgb255=(195,215,223)
		rgb255=(213,231,236)
		rgb255=(219,231,234)
		rgb255=(218,216,218)
		rgb255=(217,204,206)
		rgb255=(216,188,190)
		rgb255=(215,176,178)
		rgb255=(215,163,166)
		rgb255=(214,147,150)
		rgb255=(213,135,138)
		rgb255=(212,120,122)
		rgb255=(211,107,110)
		rgb255=(210,92,94)
		rgb255=(209,79,82)
		rgb255=(208,64,66)
		rgb255=(207,51,54)
		rgb255=(207,40,42)
	},
	colormap={natBlueRedA30}{
		rgb255=(0,39,82)
		rgb255=(18,55,95)
		rgb255=(36,71,108)
		rgb255=(60,93,125)
		rgb255=(78,110,138)
		rgb255=(102,131,156)
		rgb255=(120,148,169)
		rgb255=(145,170,186)
		rgb255=(163,186,199)
		rgb255=(187,208,217)
		rgb255=(205,224,230)
		rgb255=(219,236,239)
		rgb255=(219,223,226)
		rgb255=(218,213,216)
		rgb255=(217,201,204)
		rgb255=(216,191,194)
		rgb255=(216,179,181)
		rgb255=(215,169,172)
		rgb255=(214,157,159)
		rgb255=(214,147,150)
		rgb255=(213,138,140)
		rgb255=(212,125,127)
		rgb255=(211,115,118)
		rgb255=(211,103,105)
		rgb255=(210,93,96)
		rgb255=(209,81,83)
		rgb255=(209,71,73)
		rgb255=(208,58,61)
		rgb255=(207,49,51)
		rgb255=(207,40,42)
	},
	colormap={natBlueRedB30}{
		rgb255=(207,40,42)
		rgb255=(209,82,84)
		rgb255=(212,125,127)
		rgb255=(216,182,184)
		rgb255=(219,224,227)
		rgb255=(212,230,235)
		rgb255=(204,223,229)
		rgb255=(193,214,222)
		rgb255=(186,207,216)
		rgb255=(175,198,209)
		rgb255=(168,191,203)
		rgb255=(160,184,197)
		rgb255=(150,174,190)
		rgb255=(142,167,184)
		rgb255=(131,158,177)
		rgb255=(124,151,171)
		rgb255=(113,141,164)
		rgb255=(106,134,158)
		rgb255=(95,125,151)
		rgb255=(87,118,145)
		rgb255=(80,111,139)
		rgb255=(69,102,132)
		rgb255=(62,95,126)
		rgb255=(51,85,119)
		rgb255=(43,78,113)
		rgb255=(33,69,106)
		rgb255=(25,62,100)
		rgb255=(15,53,93)
		rgb255=(7,46,87)
		rgb255=(0,39,82)
	},
	colormap={natBlueRedC30}{
		rgb255=(0,39,82)
		rgb255=(9,47,88)
		rgb255=(19,56,95)
		rgb255=(32,68,105)
		rgb255=(41,76,112)
		rgb255=(54,88,121)
		rgb255=(64,97,128)
		rgb255=(76,108,137)
		rgb255=(86,117,144)
		rgb255=(99,128,153)
		rgb255=(109,137,160)
		rgb255=(118,146,167)
		rgb255=(131,157,177)
		rgb255=(141,166,183)
		rgb255=(153,178,193)
		rgb255=(163,186,200)
		rgb255=(176,198,209)
		rgb255=(186,207,216)
		rgb255=(198,218,225)
		rgb255=(208,227,232)
		rgb255=(218,236,239)
		rgb255=(218,215,218)
		rgb255=(217,196,199)
		rgb255=(215,170,172)
		rgb255=(214,150,153)
		rgb255=(212,124,127)
		rgb255=(211,105,107)
		rgb255=(209,79,81)
		rgb255=(208,59,61)
		rgb255=(207,40,42)
	},
	colormap={natBlueRedD30}{
		rgb255=(0,39,82)
		rgb255=(25,62,100)
		rgb255=(51,85,119)
		rgb255=(86,117,144)
		rgb255=(112,140,163)
		rgb255=(147,172,188)
		rgb255=(173,195,207)
		rgb255=(207,226,232)
		rgb255=(219,233,236)
		rgb255=(219,222,225)
		rgb255=(218,214,217)
		rgb255=(217,206,209)
		rgb255=(217,196,198)
		rgb255=(216,187,190)
		rgb255=(216,177,179)
		rgb255=(215,169,171)
		rgb255=(214,158,160)
		rgb255=(214,150,152)
		rgb255=(213,139,142)
		rgb255=(213,131,133)
		rgb255=(212,123,125)
		rgb255=(211,112,115)
		rgb255=(211,104,106)
		rgb255=(210,93,96)
		rgb255=(210,85,87)
		rgb255=(209,74,77)
		rgb255=(208,66,69)
		rgb255=(208,56,58)
		rgb255=(207,48,50)
		rgb255=(207,40,42)
	},
	colormap={natBlueRedE30}{
		rgb255=(0,39,82)
		rgb255=(17,54,94)
		rgb255=(34,70,107)
		rgb255=(57,91,123)
		rgb255=(75,106,136)
		rgb255=(98,127,152)
		rgb255=(115,143,165)
		rgb255=(138,164,182)
		rgb255=(156,180,194)
		rgb255=(179,201,211)
		rgb255=(196,216,223)
		rgb255=(213,232,236)
		rgb255=(219,228,231)
		rgb255=(218,218,221)
		rgb255=(217,205,208)
		rgb255=(217,195,198)
		rgb255=(216,182,185)
		rgb255=(215,173,175)
		rgb255=(214,160,162)
		rgb255=(214,150,153)
		rgb255=(213,140,143)
		rgb255=(212,127,130)
		rgb255=(212,117,120)
		rgb255=(211,104,107)
		rgb255=(210,95,97)
		rgb255=(209,82,84)
		rgb255=(209,72,74)
		rgb255=(208,59,61)
		rgb255=(207,49,51)
		rgb255=(207,40,42)
	},
	}
\pgfplotsset{compat=1.14}
\pgfplotsset{minor grid style={dashed}}
\pgfplotsset{
	cycle list/.define={linBR}{[of colormap=linBRmain]},
	colormap={linBR}{
		indices of colormap=
		(0,2,...,\pgfplotscolormaplastindexof{linBRmain} of linBRmain)
	}, 
}%;
\pgfplotsset{
	cycle list/.define={linBR}{[of colormap=linBRmain]},
	colormap={linBRrev}{
		indices of colormap=(
			\pgfplotscolormaplastindexof{linBRmain},...,0 of linBRmain
		)
	},
	cycle list/.define={linBRrev}{[of colormap=linBRrev]},
}%;
\pgfplotsset{
	colormap={linBR7rev}{
		indices of colormap=(
			0,3,...,\pgfplotscolormaplastindexof{linBRrev} of linBRrev
		)
	},
	cycle list/.define={linBR7rev}{[of colormap=linBR7rev]},
}
\pgfplotsset{
	colormap={linBR7}{
		indices of colormap=(
			0,3,...,\pgfplotscolormaplastindexof{linBRmain} of linBRmain
		)
	},
	cycle list/.define={linBR7}{[of colormap=linBR7]},
}
\pgfplotsset{
	colormap={linBR6rev}{
		indices of colormap=(
			0,3,...,\pgfplotscolormaplastindexof{linBRrev} of linBRrev
		)
	},
	cycle list/.define={linBR6rev}{[of colormap=linBR6rev]},
}
\pgfplotsset{
	colormap={linBR6}{
		indices of colormap=(
			0,3,...,\pgfplotscolormaplastindexof{linBRmain} of linBRmain
		)
	},
	cycle list/.define={linBR6}{[of colormap=linBR6]},%
}

\hypersetup{
	colorlinks = true,
	linkcolor = azul-gris,
	citecolor = naranja_1,
	urlcolor = orange,
	runcolor = black,
	breaklinks = true,
}

\newcommand{\myName}{A. Toral-Lopez}

\titleformat{name=\chapter, numberless}											% Command
[display]																									% Shape
{\bfseries\large}																					% Format
{%\color{blue} \linethickness{0.01mm} \line(1,0){350}			% Label
}
{-7ex}																										% Horizontal Separation
{\centering\MakeUppercase}
[\vspace{5ex}]					
\titlespacing*{name=\chapter, numberless}{0pt}{120pt}{6pt}

\ifdefined\SingleRep
	\titleformat{\chapter}[display]
	{\centering \bfseries \Large} %{\huge}
	{}%\filleft\MakeUppercase{\chaptertitlename}}
	{0ex}
	{}%{Report }
	[\vspace{0.1cm} \centering \large \sf \myName \vspace{0.2cm} \rule{1.0\textwidth}{0.5pt}] % [\LARGE]
	\titlespacing*{\chapter}{0pt}{-20pt}{12pt}
\else
	\titleformat{\chapter}[display]
	{\centering \bfseries \huge}
	{}%\filleft\MakeUppercase{\chaptertitlename}}
	{0ex}
	{}
	[\rule{1.0\textwidth}{0.5pt}]
	\titlespacing*{\chapter}{0pt}{-20pt}{12pt}
\fi

\fancypagestyle{plain}{
	\fancyhead{}
	
}
\newcommand{\secref}[1]{\hyperref[#1]{Section \ref*{#1}}}
\newcommand{\appref}[1]{\hyperref[#1]{Appendix \ref*{#1}}}
\newcommand{\chref}[1]{\hyperref[#1]{Chapter \ref*{#1}}}
\newcommand{\figref}[1]{\hyperref[#1]{Figure \ref*{#1}}}
\newcommand{\sfigref}[2]{\hyperref[#1]{Figure \ref*{#1}#2}}
\newcommand{\nfigref}[2]{\hyperref[#1]{\ref*{#1}#2}}
\newcommand{\tabref}[1]{\hyperref[#1]{Table \ref*{#1}}}

\newcommand{\expFnc}[1]{\mathlarger{\mathlarger{e^{#1}}}}

\DeclareUnicodeCharacter{2212}{-}

\tikzstyle{textFil} = [text opacity=1, fill opacity=0.6]
\tikzstyle{startstop} = [rectangle, rounded corners, minimum width=3cm, minimum height=1cm,text centered, draw=black, fill=red!30]
\tikzstyle{io} = [trapezium, trapezium left angle=70, trapezium right angle=110, minimum width=3cm, minimum height=1cm, text centered, draw=black, fill=blue!30]
\tikzstyle{process} = [rectangle, minimum width=3cm, minimum height=1cm, text centered, draw=black, fill=orange!30]
\tikzstyle{decision} = [diamond, minimum width=3cm, minimum height=1cm, text centered, draw=black, fill=green!30]
\tikzstyle{arrow} = [thick, ->, >=stealth]

\pgfkeys{/tikz/savenumber/.code 2 args={\global\edef#1{#2}}}

\newcolumntype{L}[1]{>{\raggedright\arraybackslash}p{#1}}
\newcolumntype{C}[1]{>{\centering\arraybackslash}p{#1}}
\newcolumntype{R}[1]{>{\raggedleft\arraybackslash}p{#1}}
\makeatletter
\def\clearpage{%
	\ifvmode
	\ifnum \@dbltopnum =\m@ne
	\ifdim \pagetotal <\topskip
	\hbox{}
	\fi
	\fi
	\fi
	\newpage
	\thispagestyle{empty}
	\write\m@ne{}
	\vbox{}
	\penalty -\@Mi
}
\makeatother

\pgfplotsset{legend style={fill opacity=0.7, text opacity=1, draw=none}}

\begin{document}
\preprint{APS/123-QED}

\title{Hopping transport regimes and dimensionality transition: a unified Monte Carlo Random Resistor Network approach}

\author{Alejandro Toral-Lopez}
\affiliation{Dipartimento di Ingneneria della Informazione, University of Pisa, Via Caruso 16, 56126, Pisa, Italy}
\email{alejandro.lopez@ing.unipi.it}
\author{Damiano Marian}
\affiliation{Dipartimento di Fisica “E. Fermi”, University of Pisa, Largo Pontecorvo 3, 56127 Pisa, Italy}
\author{Gianluca Fiori}
\affiliation{Dipartimento di Ingneneria della Informazione, University of Pisa, Via Caruso 16, 56126, Pisa, Italy}%

\date{\today}

\begin{abstract}
Hopping transport, characterized by carrier tunneling between localized states, is a key mechanism in disordered materials such as organic semiconductors, perovskites, nitride alloys, and 2D material-based inks. Two main regimes are typically observed: Variable Range Hopping and Nearest Neighbor Hopping, with a transition between them upon temperature variation. Despite numerous experimental observations, the modeling of this transition remain insufficiently explored and not fully understood. In this work, we present an in-house Monte Carlo Random Resistor Network-based simulator capable of capturing both hopping transport regimes. We demonstrate how material properties that define the network, such as localization length and the spatial and energetic distribution of sites, determine the dominant transport regime. The simulator has been successfully validated against experimental data, showing excellent agreement, reproducing the transition from one regime to the other and accurately capturing 1D, 2D and 3D variable range hopping behavior, providing both a theoretical framework for interpreting experiments and a powerful tool for studying transport mechanisms.

\end{abstract}
\maketitle
\section{Introduction}

Hopping transport has been a subject of research since the seminal study on conductivity in doped silicon at low temperatures \cite{Shklovskii1984}. This type of transport is characterized by the tunneling of carriers between available sites in the material, resembling a hopping mechanism. One of the primary requirement for observing this phenomenon is the presence of localized states in the material, typically arranged in a disordered manner. In recent years, hopping transport has gained increasing interest, as it has been observed in various physical systems, including amorphous materials such as organic semiconductors \cite{Yi2016, Yamashita2014}, perovskites \cite{Javed2023} or nitride alloys \cite{Godet2007}. Two-dimensional (2D) systems have attracted increasing interest in the field of electronics in the last years, and amorphous systems can also be found within this class. While hopping transport is a well-known mechanism in 2D materials themselves \cite{Qiu2013}, it has also been observed in 2D silica glass \cite{Huang2013}, amorphous BN \cite{Hong2020}, amorphous graphene \cite{Toh2020, Tian2023}, amorphous C-N \cite{Bai2024}, and in inks based on 2D materials (e.g., graphene, TMDs, and MXenes) \cite{Piatti2021, Mondal2021, Gabbett2024, Grillo2025}. In particular, in the context of 2DM ink-based devices, hopping transport has attracted significant attention. The transport properties depend on the material itself, the ink preparation and fabrication processes, as well as device fabrication \cite{Piatti2021, Mondal2021}.

Usually, two distinct transport regimes can be identified: Nearest Neighbor Hopping (NNH) \cite{Shklovskii1984} and Variable Range Hopping (VRH) \cite{Mott_1969, Shklovskii1984}. NNH was among the first models used to analyze charge transport in these types of materials, with the aim to explain their thermally activated behavior. In this regime, charge exchange occurs between sites that are close in both space and energy, resulting in an activated transport process upon temperature variation. On the other hand, VRH, firstly introduced by N.~F.~Mott \cite{Mott_1969}, describes a more complex mechanism in which the hopping distance varies both spatially and energetically in order to optimize the hopping probability. These two regimes can be directly extracted from experimental data by analyzing the linearity of the conductivity on a logarithmic scale as a function of $T^{-1}$ (NNH) and $T^{-1/(d+1)}$ (VRH), where $d$ represents the dimensionality of the system and $T$ the temperature. Is worth mentioning that there is an additional type of VRH, the Efros–Shklovskii VRH (ES VRH). It is observed when Coulomb interactions between the sites are present and depicts a $T^{-1/2}$ temperature dependency, independent of the dimensionality of the system. In the present article, we will only focus on the Mott-VRH transport. Experimentally, a transition between these two transport regimes is commonly observed \cite{Sun2020, Rudra2021, Piatti2021, Mondal2021, Aghashahi2022, Lee2024}, with VRH dominating at lower temperatures and a shift to NNH as the temperature increases. This transition to the NNH regime is generally attributed to a certain activation energy $\Delta E_{\rm{NNH}}$ of NNH conduction, which becomes accessible once the temperature reaches $k_{\rm{B}}T \sim \Delta E_{\rm{NNH}}$, with $k_{\rm{B}}$ the Boltzmann constant. Currently, most theoretical and simulation-based approaches focus either on NNH or VRH \cite{Ortuno2021}, or on rationalizing experimental data by extracting an effective mobility that can be used in semiclassical drift-diffusion transport simulations \cite{Nenashev2013, Upreti2019}. However, to the best of our knowledge, no study has directly addressed the VRH-NNH transition or how it is influenced by the material properties.

In this work, we address this gap by using an in-house Random Resistor Network (RRN) based simulator. Other methods, such as Monte Carlo approaches for single - electron transitions \cite{Bhandari2023, Upreti2019} or percolation-based models \cite{Li2024}, are more computationally demanding when applied to large systems, and are generally better suited for analyzing individual hopping paths rather than the macroscopic conductivity. In this context, the Resistor Random Network (RRN) approach is a more suitable option for evaluating the conductivity of the system, enabling flexibility in controlling the spatial and energetic distribution of hopping sites. When combined with a Monte Carlo scheme, this method allows the evaluation of hopping conductivity across different simulation scenarios. We would like to emphasize that in the present study we do not consider time-dependent effects, as our focus is mainly on the impact of spatial and energetic disorder on the conductivity of the system. The analysis of disorder from a temporal perspective requires different, but complementary,  theoretical frameworks, such as the Continuous Time Random Walk (CTRW) model \cite{Shlesinger2017}. In the latter, transport is described at the particle level, accounting for the stochastic dynamics of hopping distances and site occupation times. Thus, our approach, as well as those mentioned above, should be considered complementary, since they can provide different information on the system.

This approach allows us to systematically investigate the interplay of material properties, such as localization length and density of states (DoS), with the observed transport regimes. To our knowledge, this is the first study that uses detailed numerical simulations to elucidate the factors governing the VRH–NNH transition, while providing new insights into the effect of material properties on transport behavior. This unique perspective paves the way for a deeper understanding of hopping transport mechanisms and their optimization for practical applications. The methodology and tools developed here can be highly useful in the study of 2D material-based inks and also in emerging memristive devices, where disorder plays a crucial role and VRH transport mechanisms have been recently observed~\cite{Wang2023}.

\section{Simulation methods}

\subsection{The Random-Resistor Network model.}
 
\begin{figure}[h!!!]
    \centering
    \includegraphics[width=0.95\textwidth]{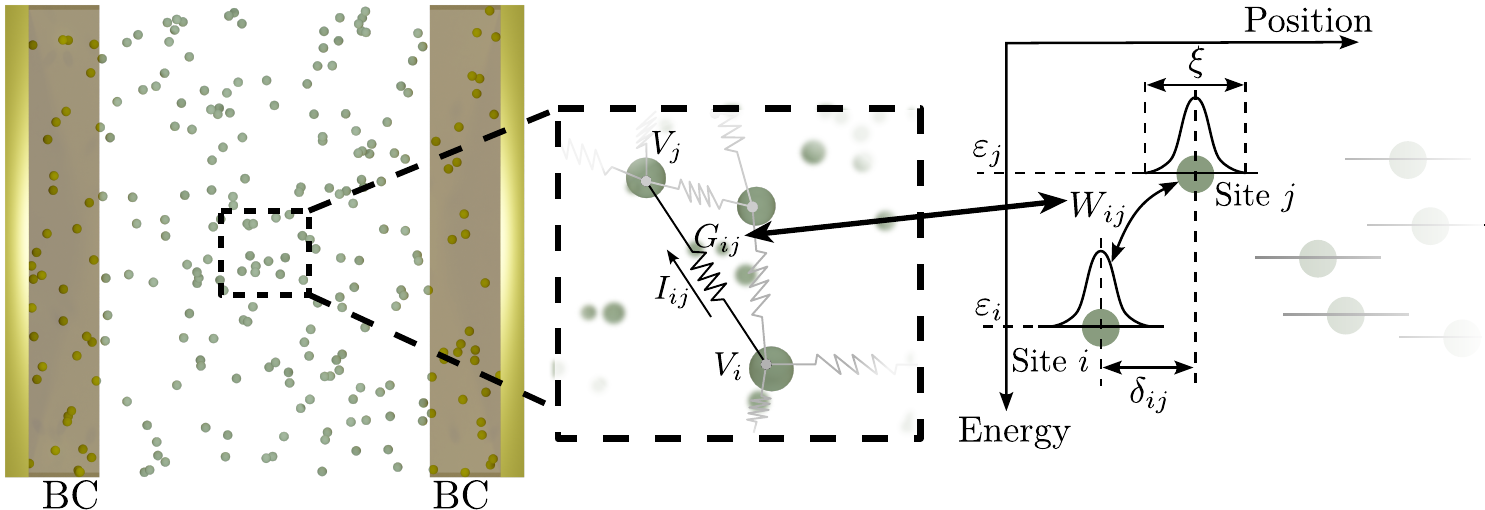}
    \caption{Sketch of the resistor network to model the system with different sites. In the RRN model, the connection between each pair of nodes is characterized by a conductance $G_{ij}$ and an associated current $I_{ij}$, which depends on that conductivity and the potential of each site $V_i$. Dirichlet boundary conditions for the potential are defined in correspondence of contact potential.}
    \label{fig:netRRNscheme}
\end{figure}

Amorphous materials, such as organic semiconductors, printed devices, and others, share a common feature: charge transport in these disordered systems typically does not occur through free carriers, but rather via hopping between localized states. For this reason, we model the active region of the material as a cloud of available sites, as schematically depicted in the left panel of \figref{fig:netRRNscheme}. The hopping of an electron from one site to another occurs over time; however, one can define an effective transition rate between sites, $W_{ij}$, that captures the average hopping behavior under equilibrium conditions. Assuming low field conditions, it is possible to define a conductance associated with $W_{ij}$~\cite{Shklovskii1984, Tessler2009, Ortuno2021}:

\begin{equation}
    G_{ij} = \frac{q^2}{k_{\rm B}T}W_{ij} = \frac{q^2v_0}{k_{\rm B}T}\expFnc{-2\frac{\delta_{ij}}{\xi}}\expFnc{-\frac{\varepsilon_{ij}}{k_{\rm B}T}}
    \label{eq:defGij}
\end{equation}
where $q$ is the electron charge, $\xi$ the localization length, $k_{\rm B}$ the Boltzmann constant, $T$ the temperature, $v_0$ the average transition rate, $\delta_{ij}$ the spatial distance between $i$-$j$ sites and $\varepsilon_{ij}$ the energetic distance, which is defined as \cite{Shklovskii1984}:

\begin{equation}
    \varepsilon_{ij} = \frac{1}{2}\left(|\varepsilon_{i}| + |\varepsilon_{j}| + |\varepsilon_{i} - \varepsilon_{j}|\right) = \frac{1}{2}\left(|\varepsilon_{i}| + |\varepsilon_{j}| + |\delta\varepsilon_{ij}|\right),
    \label{eq:difenerg}
\end{equation}
% which in this case is considered as the reference level (i.e., $E_{\rm F} = 0$~eV)
where $\varepsilon_{i}$ is the energy associated to the $i$-th site with respect to the Fermi level. The cloud of sites can be modeled as a network of conductivities/resistances, as shown in the central and right panels of \figref{fig:netRRNscheme}, whose values depend on the spatial and energetic position of the sites. In general terms, one has a set of $N$ nodes, each with a specific spatial position and a given energy. The position can be defined either randomly or according to a specific grid, while the energy distribution is usually set randomly according to a specific probability density function (PDF) related to the energetic Density of States (DoS) profile associated to the sites.

Given a network configuration, usually, the main objective is to obtain the total conductance between two contacts that defines the boundary of the network. A possible approach to address this problem is to solve the RRN \cite{Ortuno2021,Tessler2009}, evaluating the total conductance of the system by assuming a test bias and solving the Kirchhoff equations in all the nodes, and to obtain the total current through the network \cite{Ortuno2021}. Then, given the $i$-th node, the total current reads:

\begin{equation}
    \sum_{j \neq i}{I_{ij}} = \sum_{j \neq i}{\left(V_i - V_j\right)G_{ij}} = 0, 
\end{equation}
where, $I_{ij}$ is the current between nodes $i$ and $j$, $G_{ij}$ the conductance, and $V_{i}$ the potential at node $i$. This equation is used to compute the potential at each node, except in the regions where boundary conditions (BCs) are defined (see the darker areas near the contacts in \figref{fig:netRRNscheme}). In those regions, the potential is fixed according to the values specified by the corresponding BCs. Once the potential at each node has been obtained, it is possible to calculate the current flowing through the network. If only two contacts are considered, the input current $(I_{\rm in})$ can be obtained by combining the current of all nodes connected to one contact, so to evaluate the total conductivity $\sigma$:

\begin{equation}
    I_{\rm in} = \sum_{k\,\,{\rm in}\,\,BC_{\rm in}}{\sum_{j\neq k}I_{kj}} \Rightarrow \sigma = \left[\frac{V_{\rm in} - V_{\rm out}}{I_{\rm in}}\right].
    \label{eq:condtot}
\end{equation}

Once the RRN is solved over a range of temperatures and the total conductivity $(\sigma)$ as a function of $T$ is obtained, one can directly evaluate its temperature dependence and determine the dominant hopping transport mechanism, i.e., NNH, VRH, or a combination of both in different temperature ranges.

% ==========================================================================

In the following, the presented model is used to analyze the influence of material parameters, such as localization length and density of states, on the transport behavior. This is done by performing numerical experiments in order to rationalize and explore the VRH-NNH transition. However, directly analyzing the conductivity in Eq.~\eqref{eq:condtot} is rather challenging due to the complexity of the expression, which involves a non-trivial combination of parallel and series conductances when considering $N$ sites randomly distributed both energetically and spatially. Due to this complexity, we adopt a step-by-step analysis. We begin by focusing on a simplified two-node network, which allows for a more tractable investigation. This simplified case provides valuable insights into the mechanisms at play and serves as a guide for subsequent RRN simulations involving larger and more realistic networks.

\subsection{The two-sites model (TsM)}

As previously mentioned, a two-node network is sufficiently simple to offer valuable insights into the NNH and VRH transport regimes. The model, by itself, is too simple to fully explain experimental data, but it is sufficient to identify the combinations of parameters that lead to each transport regime, serving as a useful reference for setting up calculations for the full network. In the following, we refer to this as the two-site model (TsM). The main objective of this analysis is to demonstrate how Eq.~\eqref{eq:defGij} captures both the $T^{-1}$ (NNH) and $T^{-1/(d+1)}$ (VRH) trends, and the regions in the space of parameters where each of them are expressed. Specifically, by computing $\ln(G_{ij})$, we obtain:

\begin{equation}
        \ln(G_{ij}) = \ln\left(\frac{q^2v_0}{k_{\rm B}T}\right) - \frac{2\delta_{ij}}{\xi} - \left(\frac{\varepsilon_{ij}}{k_{\rm B}}\right)T^{-1}.
    \label{eq:Gnnh}
\end{equation}
Eq.~\eqref{eq:Gnnh} clearly shows a linear dependence on $T^{-1}$ in the last term, indicating NNH hopping transport behavior. In this case, $G_{ij}$ corresponds to NNH transport in the sense that maximum conductivity is achieved when both the physical ($\delta_{ij}$) and energetic ($\varepsilon_{ij}$) distances are minimal. It is also worth noting that a temperature dependence appears in the first term of Eq.~\eqref{eq:Gnnh}; however, its role will be discussed at the end of this section.

Now we show that Eq.~\eqref{eq:defGij} can also describe VRH transport, but this occurs only for specific values of the material parameters, i.e., the localization length $\xi$, the density of states (DoS), and the density of nodes.  Following Ref.~\cite{Tessler2009}, if we consider a spatial region $\Omega_{\rm d}$ around a given node, a hop with energy $\varepsilon_{ij} = \Delta \varepsilon$ is allowed only if:

\begin{equation}
    \Omega_{\rm d}\Delta \varepsilon\rho = 1 \;\;\;\;\; \Rightarrow \;\;\;\;\;  \Delta \varepsilon = \frac{1}{\Omega_{\rm d}\rho}
    \label{eq:vrher}
\end{equation}
where, $\rho$ is the density of states (DoS) of the system, and $\Omega_{\rm d} = \Omega_0 R^d$, where $d$ is the dimensionality of the system. The constant $\Omega_0$ depends on the dimensionality: $\Omega_0 = \frac{4}{3}\pi$ for $d = 3$, $\Omega_0 = \pi$ for $d = 2$, and $\Omega_0 = 1$ for $d = 1$. $R$ represents the hopping range, which, in the case of the TsM, corresponds to the distance between nodes, $\delta_{ij}$. Eq.~\eqref{eq:vrher} therefore establishes a specific relation between $\varepsilon_{ij}$ and $\delta_{ij}$. Substituting this relation into Eq.~\eqref{eq:defGij} allows one to determine the value of $\delta_{\rm opt}$ (or, equivalently, $\varepsilon_{\rm opt}$) that maximizes the conductance $G_{ij}$:

\begin{equation}
    G_{ij} = G_0\expFnc{f(\delta_{ij})} \;\;\; \text{with} \;\;\; f(\delta_{ij}) = \frac{2\delta_{ij}}{\xi} + \frac{\delta_{ij}^{-d}}{k_{\rm B}T\Omega_0\rho} 
    \label{eq:Gopt}
    \end{equation}
\begin{equation} 
    \frac{\partial f(\delta_{ij})}{\partial \delta_{ij}} = 0 \;\;\; \Rightarrow \;\;\; \left\lbrace\begin{array}{c}
        \delta_{\rm opt} = \left(\frac{d\xi}{2\Omega_0 k_{\rm B}\rho}T^{-1}\right)^\frac{1}{d+1} \\
        \varepsilon_{\rm opt} = \left(\frac{1}{\Omega \rho}\left(\frac{2}{d}\frac{k_{\rm B}T}{\xi}\right)^d\right)^\frac{1}{d+1} 
    \end{array}
    \right.
\label{eq:dopt}            
\end{equation}
If we substitute Eq.~\eqref{eq:dopt} and Eq.~\eqref{eq:vrher} in Eq.~\eqref{eq:Gnnh}, we obtain: 

\begin{equation}
\ln(G_{ij}) = \ln\left(\frac{q^2v_0}{k_{\rm B}T}\right) - \frac{2\delta_{\rm opt}}{\xi} + \frac{\delta_{\rm opt}^{-d}}{k_{\rm B}T\Omega_0\rho} = \ln\left(\frac{q^2v_0}{k_{\rm B}T}\right) - \left(\frac{T_0}{T}\right)^\frac{1}{d+1}
\label{eq:Gvrh}
\end{equation}
where $T_0 = \left(\frac{(d+1)^{d+1}}{\Omega_0 k_{\rm B}\rho}\left(\frac{2}{d\xi}\right)^{d}\right)$. Eq.~\eqref{eq:Gvrh} shows a linear dependence on $T^{-\frac{1}{d+1}}$ in the last term, as expected for VRH transport. As for Eq.~\eqref{eq:Gnnh}, we also observe a temperature dependence in the first term of Eq.~\eqref{eq:Gvrh}, which will be discussed at the end of this section.

Let's analyze in detail the relations we have just derived. Eq.~\eqref{eq:dopt} shows that the optimum hopping distance ($\delta_{\rm opt}$) depends not only on the material parameters ($\xi$ and $\rho$), but also on the temperature. Also, $\varepsilon_{\rm opt}$ depends on $\xi$, $\rho$ and $T$. Therefore, for a given system configuration, $\delta_{\rm opt}$ and $\varepsilon_{\rm opt}$ are temperature-dependent. In particular, as indicated by Eq.~\eqref{eq:dopt}, the optimum physical hopping distance is expected to increase as the temperature decreases, while the optimum energy distance decreases. This implies that VRH transport, due to its temperature dependence, can only be observed within a certain temperature range, which is determined by the characteristics of the network and the material parameters, summarized in the characteristic temperature $T_0$.

\begin{figure}[th]
    \centering
    \includegraphics[width=0.7\textwidth]{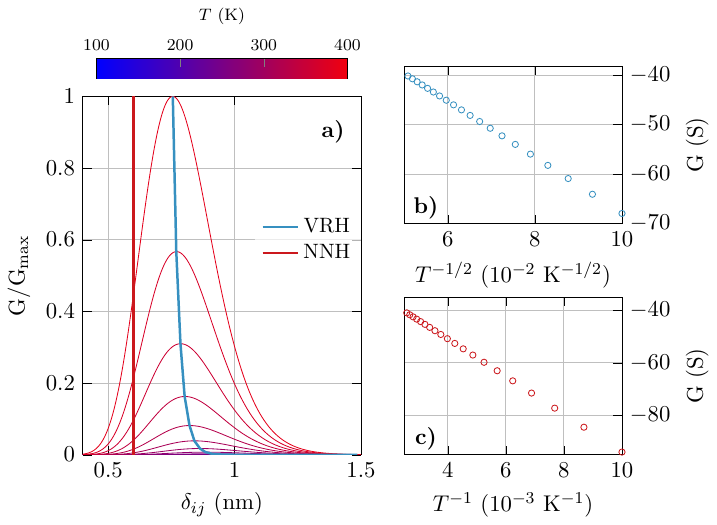}
    \caption{(a) Normalized conductance of the TsM as a function of the distance between the two nodes $\delta_{ij}$. Blue curve shows the profile of $\delta_{\rm opt}$ while red curve indicate the profile for a fixed distance between nodes $\delta_{ij}$. (b) Conductance, in the logarithmic scale, of the blue curve in (a) vs $T^{-1/2}$: the linearity clearly indicates it follows the VRH. (c) Conductance, in logarithmic scale, of the red curve in (a) vs $T^{-1}$: the linearity clearly indicates it follows the NNH.}
    \label{fig:optcond1d1nm}
\end{figure}

In \sfigref{fig:optcond1d1nm}{a}, the conductance $G_{ij}$ as expressed in Eq.~\eqref{eq:Gopt} is plotted for a one-dimensional system ($d = 1$) and for several temperatures, as a function of the hopping distance. The blue curve in the figure indicates the profile of $\delta_{\rm opt}$ at each temperature, obtained from Eq.~\eqref{eq:dopt}. As expected, it follows the maximum of the corresponding $G_{ij}$ profiles. When the conductance profile along this blue line is plotted on a logarithmic scale (\sfigref{fig:optcond1d1nm}{b}) in the $T^{-1/2}$-scale, it exhibits the linear trend characteristic of VRH.

Conversely, if we fix the distance between nodes, represented by the red line in \sfigref{fig:optcond1d1nm}{a}, a linear trend emerges only when the conductivity is plotted against $T^{-1}$, as shown in \sfigref{fig:optcond1d1nm}{c}. This behavior is representative of the NNH regime, as it mimics a scenario in which hopping occurs only between sites separated by a fixed distance.

Thus, we have shown that, given the conductance between two nodes in Eq.~\eqref{eq:defGij}, it is possible to recover both NNH and VRH behavior depending on the material parameters and their relation, by analyzing the linearity of the conductance on a logarithmic scale as a function of $T^{-1}$ and $T^{-1/(d+1)}$, respectively.

As noted above, both the NNH (Eq.~\eqref{eq:Gnnh}) and VRH (Eq.~\eqref{eq:Gvrh}) expressions contain the term $q^2 v_0 / k_{\rm B} T$. This term accounts for the electron-phonon coupling through the synthetic parameter $v_0$. The parameter $v_0$ has been treated differently in the literature, depending on the specific material or system under study. For example, it has been assumed to have a linear dependence on temperature (i.e., $v_0 \propto T$), a power-law dependence on $T$, or to be temperature-independent. In our case, to make minimal assumptions, we treat $v_0$ as a constant \cite{Ortuno2021}.

\section{Simulation results}

In the previous sections, we have introduced the main simulation framework based on the RRN model, along with a simplified two-site model (TsM), which assist in interpreting the simulation results and can serve to guide parameter selection. We now turn to the analysis of how different material parameters, such as localization length, density of states (DoS), etc..., influence the transition between VRH and NNH transport regimes. Using a one-dimensional network as a reference, we first examine the effects of imposing a minimum distance between sites and varying the width of the energy distribution associated with the localized states. Next, we explore how different energetic distribution profiles for the sites affects transport. Taking a step further, we increase the complexity of the network by investigating how the VRH regime is affected by dimensionality, i.e., by extending the system to three dimensions. Finally, in the last subsection, we present results that validate the RRN implementation against experimental measurements.

\subsection{Simulation of a regular 1D network.}

The first case considered is a uniformly spaced 1D network with an inter-site distance of $\delta_{\rm min} = 1$~nm, a value close to the inter-defect spacing reported in the literature for ink jet-printed 2D-flake materials \cite{Piatti2021}. The energy of each site is randomly assigned, following a uniform distribution with an energy span of $W_{\rm E} = 0.4~\rm{eV}$. The last parameter appearing in Eq.~\eqref{eq:defGij} is the localization length, $\xi$, which we vary in the range from 0.05~nm to 0.3~nm, i.e. values comparable to experimental data reported for MoS$_2$ ink-based devices \cite{Piatti2021}. According to Eq.~\eqref{eq:dopt}, derived from the TsM, for $\xi = 0.3$~nm we expect VRH behavior across the entire temperature range considered (100~K to 400~K), since the condition $\delta_{\rm opt} > \delta_{\rm min}$ is always satisfied. When $\xi$ is reduced to 0.15~nm, this condition is no longer fulfilled across the full temperature range, and a mixed VRH-NNH behavior is therefore expected for this value of $\xi$ and below.

To verify the predictions of the TsM, we performed 100 simulations for each value of $\xi$ to compute the $\langle \sigma(T) \rangle$ curves. These were linearly fitted on a $T^{-1/2}$ scale within the range $[T_{\rm min}, T_x]$, where $T_x$ is a variable upper limit. We then evaluated the fitting error using:

\begin{equation}
    \epsilon_{\rm VRH}(T_{x}) = \sqrt{\frac{1}{T_{x} - T_{\rm min}} \int_{T_{\rm min}}^{T_{x}} \left( \frac{g(T) - y(T)}{g(T)} \right)^2 dT},
    \label{eq:error}
\end{equation}
where $g(T)$ represents the linear fit for the VRH regime. When $\epsilon_{\rm VRH}$ exceeds a given threshold (e.g., $5 \cdot 10^{-3}$), this indicates a deviation from the VRH trend, indicating a transition to NNH behavior. The corresponding value of $T_x$ is then defined as the transition temperature, $T_{\rm c}$: below $T_{\rm c}$ the curve follows VRH behavior, while above it, NNH dominates. 

\begin{figure}[th]
    \centering
    \includegraphics[width=0.95\textwidth]{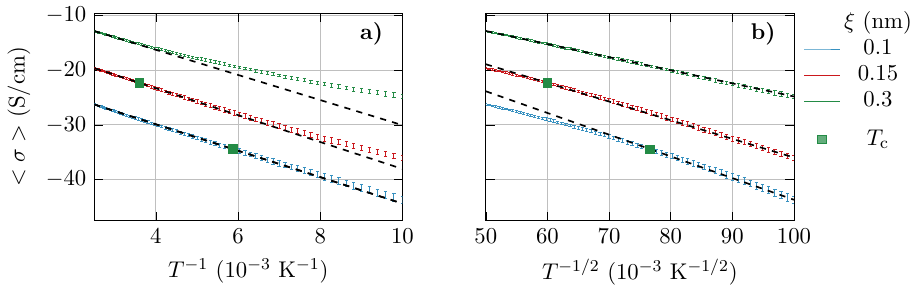}
    \caption{Conductivity curves (solid lines) for $\xi$ equal to 0.1~nm, 0.15~nm, 0.3~nm, along with their corresponding $T_{\rm c}$ (green squares), in the NNH ($T^{-1}$) and VRH ($T^{-1/2}$) scales. Dash lines indicate the linear trend for each case. The error bars indicate standard deviation with respect to the mean values.}
    \label{fig:net1d1nmsigmadata}
\end{figure}

In \figref{fig:net1d1nmsigmadata}, we report the computed conductivity in logarithmic scale, plotted as a function of both $T^{-1}$ and $T^{-1/2}$, along with the corresponding transition temperatures $T_{\rm c}$ for different values of $\xi$ (indicated by squares). Dashed lines represent the ideal trends for NNH (\sfigref{fig:net1d1nmsigmadata}{a}) and VRH (\sfigref{fig:net1d1nmsigmadata}{b}). For $\xi = 0.15$~nm and $0.1$~nm, a transition between hopping transport regimes is observed. In contrast, for $\xi = 0.3$~nm, VRH behavior persists across the entire temperature range.

\begin{figure}[th]
    \centering
    \includegraphics[width=0.9\textwidth]{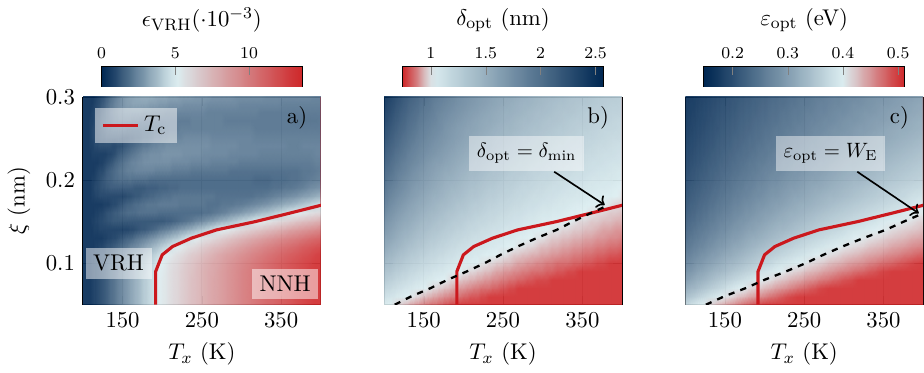}
    \caption{(a) color maps of the error for the linear fitting for VRH ($\epsilon_{\rm VRH}$) as a function of $T_x$ and $\xi$. (b, c) color maps with the optimum physical (b) and energetic (c) distances required to have a VRH behavior.}
    \label{fig:net1d1nerrmap}
\end{figure}

In \sfigref{fig:net1d1nerrmap}{a}, we present a color map of $\epsilon_{\rm VRH}$, calculated according Eq.~\eqref{eq:error}, as a function of $\xi$ and $T_{x}$, along with the $T_{\rm c}$ profile (red curve). Two distinct regions can be identified, corresponding to the VRH regime (upper-left region) and the NNH regime (lower-right corner). \sfigref{fig:net1d1nerrmap}{b} and \sfigref{fig:net1d1nerrmap}{c} show the calculated values of $\delta_{\rm opt}$ and $\varepsilon_{\rm opt}$, respectively, using the TsM for the same range of $\xi$ and $T_{x}$. In each color map, we also display the $T_{\rm c}$ curve, along with two isolines corresponding to $\delta_{\rm opt} = \delta_{\rm min}$ and $\varepsilon_{\rm opt} = W_{\rm E}$. It is worth noting that the $T_{\rm c}$ profile follows the limitations imposed by the network. In the high-temperature region and for $\xi$ values in the range of approximately 0.05–0.15~nm, we observe that it is not possible to follow the optimal hopping path (i.e., $\delta_{\rm opt} < \delta_{\rm min}$ and $\varepsilon_{\rm opt} < W_{\rm E}$). As a result, the system cannot achieve VRH behavior, and NNH transport dominates in this region.

\begin{figure}[th!]
    \centering
    \includegraphics[width=0.9\textwidth]{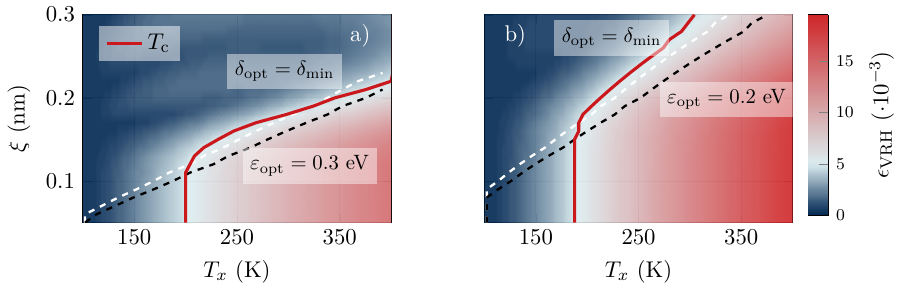}
    \caption{Color maps of $\varepsilon_{\rm VRH}$ for $W_{\rm E}$ equal to 300~meV (a) and 200~meV (b). Dashed lines indicate the $T_x$ and $\xi$ values at which $\delta_{\rm opt} = \delta_{\rm min}$ (white) and $\varepsilon_{\rm opt} = W_{\rm E}$ (black).}
    \label{fig:net1d1n200everrmap}
\end{figure}

Our analysis shows that $T_{\rm c}$ depends on the material parameters that define the network, including both the spatial and energetic distributions of the sites. To further investigate this behavior, we varied the energetic disorder by reducing the energy bandwidth $W_{\rm E}$. In \figref{fig:net1d1n200everrmap}, we present color maps of $\epsilon_{\rm VRH}$ for the same network configuration, but with $W_{\rm E} = 300$~meV (\sfigref{fig:net1d1n200everrmap}{a}) and $W_{\rm E} = 200$~meV (\sfigref{fig:net1d1n200everrmap}{b}). We observe that the temperature range in which VRH behavior dominates shifts according to the value of $W_{\rm E}$. In particular, for a fixed $\xi$, the transition temperature $T_{\rm c}$ decreases as $W_{\rm E}$ is reduced.

\subsection{Energetic profile of DoS}

\begin{figure}[th]
    \centering
    \includegraphics[width=0.8\textwidth]{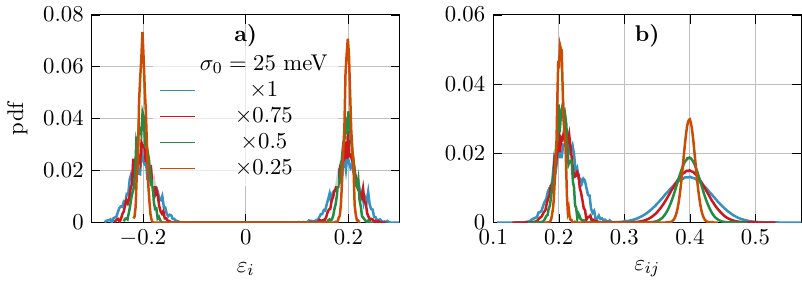}
    \caption{(a) Probability density functions used to set the energy of the nodes $\varepsilon_i$ and (b) the corresponding distribution for $\varepsilon_{ij}$. The profiles where obtained by extracting the histogram over 5000 samples of the random distribution.}
    \label{fig:tcgaussdoss}
\end{figure}

The analysis presented in the previous section assumes a constant probability density function (PDF), a common approximation in theoretical studies of amorphous materials \cite{Ortuno2021}. However, in some cases, energy states tend to localize around specific energy values \cite{Coehoorn2012, Wu2017, Zheng2019}.  In the case of 2D material-based inks, structural defects, such as chalcogen vacancies, are often present and tend to localize within a narrow energy range near the conduction and/or valence bands \cite{Sun2019}. Additionally, impurities introduced during ink preparation can result in the formation of distinct energy states.  To account for these effects, it is useful to investigate scenarios with restricted ranges of allowed energies. In such cases, energy windows between defect-related bands become inaccessible for hopping transport, introducing additional constraints that influence the optimal hopping paths in the VRH regime. To this end, we consider a PDF with double-Gaussian energy profile, as shown in \sfigref{fig:tcgaussdoss}{a}. This profile introduces variations in the range of allowed $\varepsilon_{ij}$ depending on the value of $\sigma_{\varepsilon}$, i.e. the gaussian width, as illustrated in \sfigref{fig:tcgaussdoss}{b}. Although these profiles alter the energy distribution, they do not affect the total density of states $\rho$ (see Eq.~\eqref{eq:vrher}): the number of available states for hopping within a given energy range remains unchanged, while only their distribution is modified. As a result, the relations obtained for $\delta_{\rm opt}$ and $\varepsilon_{\rm opt}$ in Eq.~\eqref{eq:dopt} remain unchanged.

\begin{figure}[th]
    \centering
    \includegraphics[width=0.95\textwidth]{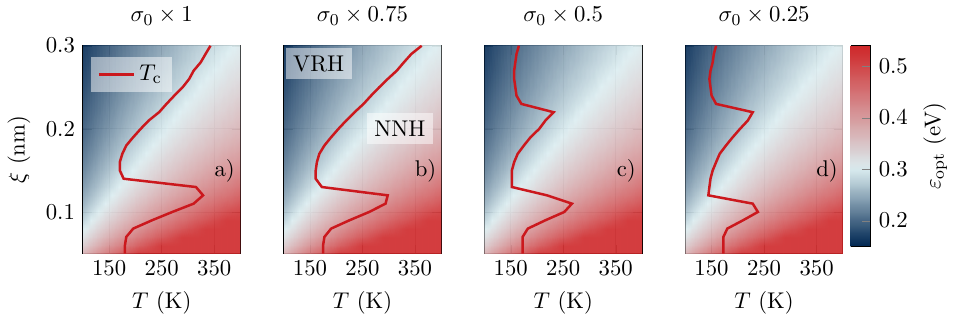}
    \caption{Energetic distance for the hopping in the optimum path (VRH) as a function of the temperature and localization length for each PDF distribution considered. The red line indicates the profile of $T_{\rm c}$ extracted for each case.}
    \label{fig:cmapgaussdosprofiledeltae}
\end{figure}

Using the profiles in \figref{fig:tcgaussdoss}{b}, we expect $T_{\rm c}$ to adapt accordingly; specifically, $T_{\rm c}$ will be lowered when the corresponding $\varepsilon_{\rm opt}$ lies outside the range of allowed $\varepsilon_{ij}$. The results confirming this behavior are shown in \figref{fig:cmapgaussdosprofiledeltae}, where $\varepsilon_{\rm opt}$ maps are displayed alongside the $T_{\rm c}$ profiles for each considered $\sigma_{\varepsilon}$. The $T_{\rm c}$ profiles for $\sigma_{\varepsilon} = \sigma_0$ (\figref{fig:cmapgaussdosprofiledeltae}{a}) and $\sigma_{\varepsilon} = 0.75 \times \sigma_0$ (\figref{fig:cmapgaussdosprofiledeltae}{b}) exhibit a minimum between $\xi = 0.15$~nm and $\xi = 0.2$~nm. This corresponds to values of $\varepsilon_{\rm opt}$ around 0.3~eV, as can be deduced from the color map (white region). As shown in \sfigref{fig:tcgaussdoss}{b}, this value falls within a region where the PDF of $\varepsilon_{ij}$ is zero, meaning VRH cannot occur, and $T_{\rm c}$ consequently remains low. When $\sigma_{\varepsilon}$ is further reduced (\figref{fig:cmapgaussdosprofiledeltae}{c} and \figref{fig:cmapgaussdosprofiledeltae}{d}), the range of forbidden $\varepsilon_{ij}$ values widens even more, resulting in low $T_{\rm c}$ for nearly all cases considered. Two peaks appear in the $T_{\rm c}$ profile, corresponding to small windows of allowed $\varepsilon_{ij}$ values. However, these windows are too narrow and $T_{\rm c}$ remains lower than 250 K. These findings highlight that the spatial distribution of nodes alone does not determine whether the system operates in the VRH regime, but the energetic distribution also plays a critical role in determining the optimal transport path within the network.

\subsection{3D network with 1D distributions}

\begin{figure}[!th]
    \centering
    \includegraphics[width=0.6\columnwidth]{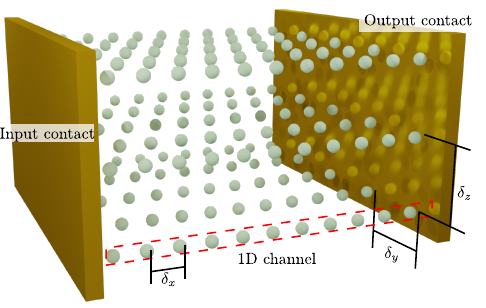}
    \caption{Structure of the 3D network where the nodes are arranged to have a preferred 1D conduction.}
    \label{fig:1dsystemnet1d}
\end{figure}

In all the previous analyses, the scenario considered was a 1D network, thus limiting the transport of the system to a single axis. The next step is to extend the study to a 3D system with a well-defined node arrangement, focusing on how a 1D-like VRH behavior moves into a fully 3D VRH regime as the localization length $\xi$ increases. Increasing $\xi$ is effectively equivalent to bringing the nodes closer together, allowing interactions along multiple spatial dimensions. To investigate this, we arrange several 1D networks in a cubic 3D space, as showed in \figref{fig:1dsystemnet1d}. In this configuration, each 1D network is repeated along the Y and Z axes, with 10 repetitions for each axis and a spacing of $5 \times \delta_x$ with $\delta_x = \delta_{\rm min}$, consistent with the parameters considered in the previous section. This setup ensures a controlled and systematic transition from 1D to 3D behavior as the $\xi$ is varied.

The aim of this analysis is twofold: first, to verify the transition from 1D to 3D VRH transport by observing how networks material parameters, in this case the localization length, affect the transport mechanism; secondly, to demonstrate the ability of the simulator to accurately reproduce and distinguish between different dimensionalities of the network. The latter could be very relevant, in cases where the input parameters, such as the localization length $\xi$, energy distribution, or node arrangement, are extracted from experimental data. This highlights the versatility of the simulator in modeling complex systems and validating experimental findings, particularly for devices where the dimensionality plays a critical role in determining transport properties.

For this analysis, we have considered a uniform energetic distribution of the sites with an energy span of $W_{\rm E} = 0.4$~eV, in order to avoid any limitations in the network due to energy profile as shown in the previous section. 

\begin{figure}[th]
    \centering
    \includegraphics[width=0.9\textwidth]{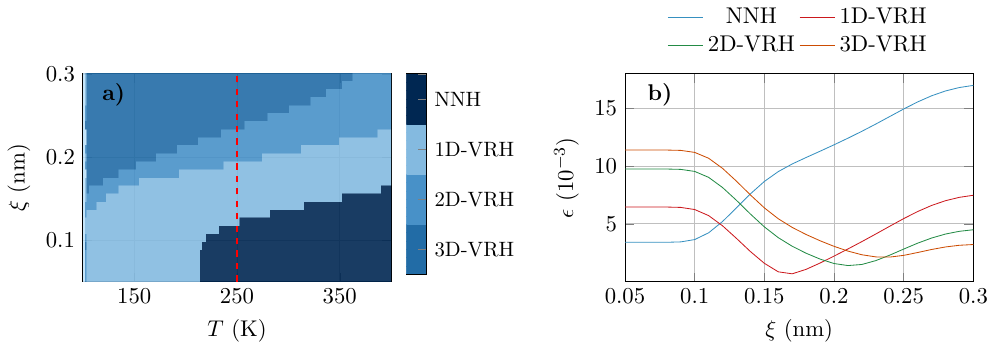}
    \caption{(a) Color maps of the regions where the network depicts 1D-VRH (light blue), 2D-VRH (blue), and 3D-VRH (dark blue). (b) Error of the linear approximation for the NNH and 1D-, 2D- and 3D-VRH transports at T = 250K as a function of the localization length $\xi$.}
    \label{fig:net3d1nerrmap}
\end{figure}

As in the previous analyses, we study the linearity of the total conductivity for the NNH ($T^{-1}$ scale) and VRH, but in the case of VRH this analysis extends to the $T^{-1/2}$, $T^{-1/3}$, and $T^{-1/4}$ scales, corresponding to 1D, 2D, and 3D VRH, respectively. In \figref{fig:net3d1nerrmap}, we report a map in the $(T, \xi)$ space (\sfigref{fig:net3d1nerrmap}{a}), highlighting the regions where each transport model yields the lowest error, indicating the most likely dominant conduction mechanism in each regime. Along with this map, \sfigref{fig:net3d1nerrmap}{b} shows how the error of the linear approximation for the NNH and VRH transports at $T=250$K.

We can observe that for temperatures higher than $T > 250$ K and localization lengths $\xi$ lower than 0.15~nm, i.e., the bottom right part of the color map, the transport is dominated by NNH, as expected from previous results on the 1D network. This is consistent with the fact that NNH is independent of the dimensionality of the system. The situation becomes particularly interesting when analyzing the VRH regime, where three distinct regions corresponding to 1D-, 2D-, and 3D-VRH can be clearly identified. In general, we observe that as the localization length increases, the optimal hopping paths for VRH is found in the whole 3D network, making this transport more favorable. Conversely, when the localization length is small, the system effectively behaves as independent arrays of nodes, and the most favorable transport regime becomes 1D VRH.

In \figref{fig:net3d1nerrmap}b), we show the fitting error from the linear approximation for each transport regime, i.e. NNH and VRH with different dimensionalities, at a fixed temperature of T = 250 K. A transition from NNH to 1D, then to 2D and finally to 3D VRH is observed as the localization length $\xi$ increases. This behavior is fully consistent with the structure of the system under investigation: a 3D network composed of well-defined, aligned 1D channels. The spacing between the nodes along the $x$ direction is five times smaller than that in the $y$ and $z$ directions. This distribution of the hopping sites imposes an ordering on the accessible paths within the network. For the lower values of $\xi$ only nearby sites participate in the transport, specifically those within the 1D channel, giving rise to a 1D-VRH trend. As the localization length increases, sites perpendicular to the channel become accessible. For a give site, the planes defined by $x-y$ and $x-z$ directions form two distinct 2D manifolds for the hopping, as they are closer than the 3D diagonal neighbours. In this regime, a 2D-VRH trend is observed. When $\xi$ increases further, sites in all directions become accessible, leading to the emergence of the 3D-VRH trend. In addition, we observe that, the localization length required for the transition between transport dimensionalities increases with temperature in accordance with the analysis reported in the previous Section for the 1D case (see \figref{fig:net1d1nerrmap}).

\subsection{Parallel network model}

The previous analysis considered a single network in which the hopping behavior evolves with temperature. However, in some experimental observations \cite{Hidaka2023, Matsuura2020,Piatti2021}, the conductivity profile exhibits a sharper change in slope that cannot be captured by a single-network model. In such cases, the transition from VRH behavior at low temperatures to a linear trend in the $T^{-1}$ scale at higher temperatures may be more accurately attributed to a shift toward band-like transport, characterized by thermal activation over a certain energy barrier \cite{Matsuura2019}, rather than to NNH. A more suitable modeling approach in this scenario considers that currents arising from different transport mechanisms flow in parallel \cite{Sun2020, Matsuura2019}.

\begin{equation} 
\sigma_{\rm tot} = \sigma_{\rm Net1} + \sigma_{\rm Net2},
\end{equation}
where, the total conductivity $\sigma_{\rm tot}$ is modeled as the sum of two contributions: $\sigma_{\rm Net1}$, associated with a network (Net1) exhibiting VRH behavior at low temperatures, and $\sigma_{\rm Net2}$, representing a second network (Net2) that dominates at higher temperatures and follows a linear trend in the $T^{-1}$ scale. In this framework, the overall transport behavior depends on the relative contributions of each network. Our simulator is capable of capturing this type of parallel transport scenario.

\begin{figure}[th]
    \centering
    \includegraphics[width=0.95\textwidth]{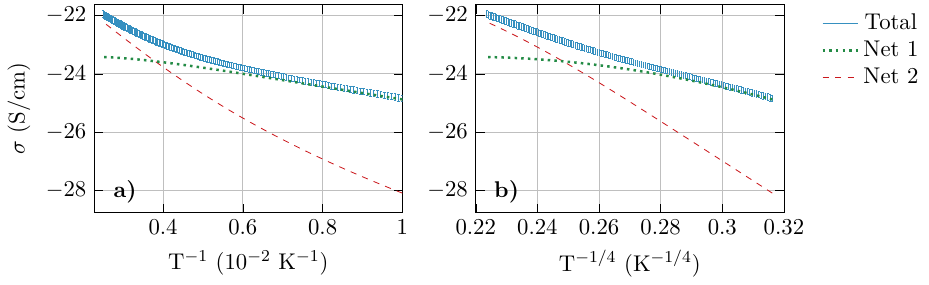}
    \caption{Conductivity of the parallel network model considering one network with regularly distributed sites (dots) and one with randomly distributed sites (dashed).}
    \label{fig:dualnet}
\end{figure}

To investigate this case, we considered two independent 3D networks with random spatial and energetic distributions of 2000 nodes. Net1 is characterized by a localization length $\xi = 0.3$~nm and an energy span $W_{\rm E} = 0.1$~eV, while Net2 has a larger energy span $W_{\rm E} = 0.4$~eV with a smaller localization length of $\xi = 0.2$~nm. For the calculations we considered 100 different configurations for each network. In \figref{fig:dualnet}, we present the average total conductivity of all these configurations, along with the corresponding average conductivity of each network. We observe that Net2 is the one with a higher temperature dependency, making Net1 to have a significant contribution in the low temperature range. Then, the resulting total conductivity profile shows a more abrupt change in the trend: a steep profile in the high temperature range as the one from Net2 that flattens as the temperature is reduced due to the contribution of Net1. Another significant change with respect to the single network modelling can be observed in \sfigref{fig:dualnet}{b}. In \sfigref{fig:net1d1nmsigmadata}{b} we observe that $\sigma$ slows is growth as $T$ is reduced, while in \sfigref{fig:dualnet}{b} $\sigma_{\rm total}$ is always increasing. This feature can be considered as a signature of the parallel network model.

\subsection{Experimental validation}

After demonstrating the capabilities of the simulator and how material parameters influence different transport regimes, particularly the transition from VRH to NNH, we now turn to the final step: validating our simulator against experimental data. Specifically, we use the experimental results from H. Matsuura et al.~\cite{Matsuura2019}, in which the authors investigate hopping transport in 4H-SiC epilayers doped with Al. In particular, they examine the transition from VRH, following a $T^{-1/4}$ dependence, to a $T^{-1}$ band-like transport regime, which closely aligns with the phenomena addressed in this work. The authors perform temperature-dependent conductivity measurements to study transport as a function of doping concentration in 4H-SiC epilayers. The dopant atoms act as hopping sites; thus, by controlling the doping concentration, the experiments offer a well-defined environment from which simulation parameters can be extracted with minimal assumptions. The doping concentration is used to define the node density which, combined with the $T_0$ value obtained from the experimental VRH fit, provides an initial estimate for the localization length $\xi$. In the case we have considered, the doping concentration corresponds to $n_0 = 1.8 \cdot 10^{20}$~cm$^{-3}$, which yields $T_0 = 18.364 \cdot 10^{6}$~K, resulting in an initial estimate of $\xi = 0.31676$~nm.

\begin{figure}[th]
    \centering
    \includegraphics[width=0.95\textwidth]{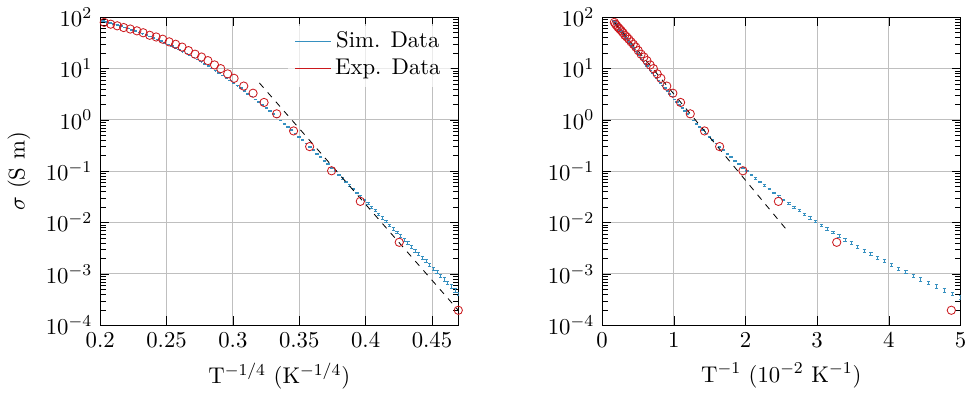}
    \caption{Experimental data (green circles) and conductivity curve (blue) obtained as the average of 100 simulations of a 3D network with a random distribution of sites, both spatially and energetically, using the parameters reported in the text. The results are shown in both the $T^{-1}$ and $T^{-1/4}$ scales, together with the experimental data points extracted from H. Matsuura et al.~\cite{Matsuura2019}. Dashed lines indicate the linear trend for each of the cases, showing that VRH trend is obtained for low temperatures (a), while NNH is obtained for high temperatures (b).}
    \label{fig:expfittingmatsuuda}
\end{figure}

Starting from the parameters previously estimated from the experimental data, we performed 100 simulations of a 3D network with a random distribution of sites, both spatially and energetically. We varied the site density $n_{\rm s}$, the energy span $W_{\rm E}$, and the localization length $\xi$ in order to reproduce the observed conductivity trends. Through this process, we found that the best agreement with the experimental data was achieved for $W_{\rm E} = 0.1$~eV, $\xi = 0.3197$~nm, and a site density of $n_{\rm s} = 0.25 n_0$. We note that the value of $\xi$ remains almost unchanged, while the reduction in site density is reasonable, as not all dopant atoms contribute to transport. The resulting fit confirms the model's ability to accurately reproduce the experimental behavior across the entire temperature range, as shown in \figref{fig:expfittingmatsuuda}.

\section{Conclusions}

In this work, we have developed an in-house simulator based on the Random Resistor Network (RRN) approach to analyze how system characteristics influence Variable Range Hopping (VRH) transport and its transition to Nearest Neighbor Hopping (NNH) upon temperature variation. In particular, we investigated how the spatial and energetic distribution of hopping sites affects the transport regime. To support this analysis, we introduced a simplified model, the Two-Sites Model (TsM), as a tool to estimate the conditions under which VRH can occur in the system. This model was then used to guide the full RRN simulations.

Our RRN simulations demonstrate that the spatial and energetic distribution of sites, along with material parameters such as the localization length, define the temperature regions where VRH transport occurs. In regimes where VRH is not favoured, the model naturally reproduce NNH behavior, or generally a $T^{-1}$ linear trend in log scale. In addition, we have shown that the simulator can accurately capture and distinguish between different dimensionalities of the network, successfully identifying 1D, 2D, and 3D VRH regimes.

Finally, we have validated our simulator against experimental data, showing excellent agreement. The simulator and the results presented here can be adapted to a wide range of scenarios, offering both a theoretical framework to interpret experimental findings and a powerful tool to investigate transport mechanisms from a theoretical perspective, like the ES VRH, where we can lever on the flexibility in the definition of the profile to distribute the energy of the nodes to introduce a ``fictitious'' Coulomb gap.

\section*{Acknoledgments}
This work was supported in part by European ERC PEP2D under Grant 770047 and through the European Union’s Horizon Europe research and innovation program (via CHIPS-JU) under the project ENERGIZE (GA 101194458). It was also supported by the Italian Ministry of Education and Research (MIUR) in the framework of the FoReLab Project (Departments of Excellence), and by European Union Next-Generation EU - National Recovery and Resilience Plan (NRRP) - MISSION 4 COMPONENT 2 INVESTMENT N 1.4 - under Grant N. CN00000013 - CUP N.I53C22000690001, in part by High Performance Computing (HPC), Big Data and Quantum Computing, Centro Nazionale 1 (CN1), Spoke 6, INVESTMENT 1.1 - under Grant N. P2022TE7Y8 - CUP I53D23000530006, 

\bibliographystyle{apsrev4-2}
\phantomsection 
\label{Bibliogra}
\bibliography{mainBib}

@Book{Shklovskii1984,
  author    = {Shklovskii, B. I. and Efros, A. L.},
  editor    = {M. Cardona, P. Fulde, HJ. Queisser},
  publisher = {Springer},
  title     = {Electronic Properties of Doped Semiconductors (Springer Series in Solid-State Sciences)},
  year      = {1984},
  isbn      = {9783540129950},
  pages     = {388},
}

@Article{Yi2016,
  author    = {Yi, H. T. and Gartstein, Y. N. and Podzorov, V.},
  journal   = {Scientific Reports},
  title     = {Charge carrier coherence and Hall effect in organic semiconductors},
  year      = {2016},
  issn      = {2045-2322},
  month     = mar,
  number    = {1},
  volume    = {6},
  doi       = {10.1038/srep23650},
  publisher = {Springer Science and Business Media LLC},
}

@Article{Yamashita2014,
  author    = {Yamashita, Yu and Tsurumi, Junto and Hinkel, Felix and Okada, Yugo and Soeda, Junshi and Zajaczkowski, Wojciech and Baumgarten, Martin and Pisula, Wojciech and Matsui, Hiroyuki and Müllen, Klaus and Takeya, Jun},
  journal   = {Advanced Materials},
  title     = {Transition Between Band and Hopping Transport in Polymer Field‐Effect Transistors},
  year      = {2014},
  issn      = {1521-4095},
  month     = oct,
  number    = {48},
  pages     = {8169--8173},
  volume    = {26},
  doi       = {10.1002/adma.201403767},
  publisher = {Wiley},
}

@Article{Javed2023,
  author    = {Javed, Muhammad and Khan, Ayaz Arif and Kazmi, Jamal and Akbar, Naeem and Khisro, Said Nasir and Dar, Amanullah and Khan Tareen, Aleem Dad and Mohamed, Mohd Ambri},
  journal   = {Journal of Rare Earths},
  title     = {Variable range hopping transport and dielectric relaxation mechanism in GdCrO3 rare-earth orthochromite perovskite},
  year      = {2023},
  issn      = {1002-0721},
  month     = jul,
  doi       = {10.1016/j.jre.2023.07.006},
  publisher = {Elsevier BV},
}

@Article{Godet2007,
  author    = {Godet, C. and Kleider, J.P. and Gudovskikh, A.S.},
  journal   = {Diamond and Related Materials},
  title     = {Frequency scaling of ac hopping transport in amorphous carbon nitride},
  year      = {2007},
  issn      = {0925-9635},
  month     = oct,
  number    = {10},
  pages     = {1799--1805},
  volume    = {16},
  doi       = {10.1016/j.diamond.2007.07.017},
  publisher = {Elsevier BV},
}

@Article{Qiu2013,
  author    = {Qiu, Hao and Xu, Tao and Wang, Zilu and Ren, Wei and Nan, Haiyan and Ni, Zhenhua and Chen, Qian and Yuan, Shijun and Miao, Feng and Song, Fengqi and Long, Gen and Shi, Yi and Sun, Litao and Wang, Jinlan and Wang, Xinran},
  journal   = {Nature Communications},
  title     = {Hopping transport through defect-induced localized states in molybdenum disulphide},
  year      = {2013},
  issn      = {2041-1723},
  month     = oct,
  number    = {1},
  volume    = {4},
  doi       = {10.1038/ncomms3642},
  publisher = {Springer Science and Business Media LLC},
}

@Article{Huang2013,
  author    = {Huang, Pinshane Y. and Kurasch, Simon and Alden, Jonathan S. and Shekhawat, Ashivni and Alemi, Alexander A. and McEuen, Paul L. and Sethna, James P. and Kaiser, Ute and Muller, David A.},
  journal   = {Science},
  title     = {Imaging Atomic Rearrangements in Two-Dimensional Silica Glass: Watching Silica’s Dance},
  year      = {2013},
  issn      = {1095-9203},
  month     = oct,
  number    = {6155},
  pages     = {224--227},
  volume    = {342},
  doi       = {10.1126/science.1242248},
  publisher = {American Association for the Advancement of Science (AAAS)},
}

@Article{Hong2020,
  author    = {Hong, Seokmo and Lee, Chang-Seok and Lee, Min-Hyun and Lee, Yeongdong and Ma, Kyung Yeol and Kim, Gwangwoo and Yoon, Seong In and Ihm, Kyuwook and Kim, Ki-Jeong and Shin, Tae Joo and Kim, Sang Won and Jeon, Eun-chae and Jeon, Hansol and Kim, Ju-Young and Lee, Hyung-Ik and Lee, Zonghoon and Antidormi, Aleandro and Roche, Stephan and Chhowalla, Manish and Shin, Hyeon-Jin and Shin, Hyeon Suk},
  journal   = {Nature},
  title     = {Ultralow-dielectric-constant amorphous boron nitride},
  year      = {2020},
  issn      = {1476-4687},
  month     = jun,
  number    = {7813},
  pages     = {511--514},
  volume    = {582},
  doi       = {10.1038/s41586-020-2375-9},
  publisher = {Springer Science and Business Media LLC},
}

@Article{Toh2020,
  author    = {Toh, Chee-Tat and Zhang, Hongji and Lin, Junhao and Mayorov, Alexander S. and Wang, Yun-Peng and Orofeo, Carlo M. and Ferry, Darim Badur and Andersen, Henrik and Kakenov, Nurbek and Guo, Zenglong and Abidi, Irfan Haider and Sims, Hunter and Suenaga, Kazu and Pantelides, Sokrates T. and Özyilmaz, Barbaros},
  journal   = {Nature},
  title     = {Synthesis and properties of free-standing monolayer amorphous carbon},
  year      = {2020},
  issn      = {1476-4687},
  month     = jan,
  number    = {7789},
  pages     = {199--203},
  volume    = {577},
  doi       = {10.1038/s41586-019-1871-2},
  publisher = {Springer Science and Business Media LLC},
}

@Article{Tian2023,
  author    = {Tian, Huifeng and Ma, Yinhang and Li, Zhenjiang and Cheng, Mouyang and Ning, Shoucong and Han, Erxun and Xu, Mingquan and Zhang, Peng-Fei and Zhao, Kexiang and Li, Ruijie and Zou, Yuting and Liao, PeiChi and Yu, Shulei and Li, Xiaomei and Wang, Jianlin and Liu, Shizhuo and Li, Yifei and Huang, Xinyu and Yao, Zhixin and Ding, Dongdong and Guo, Junjie and Huang, Yuan and Lu, Jianming and Han, Yuyan and Wang, Zhaosheng and Cheng, Zhi Gang and Liu, Junjiang and Xu, Zhi and Liu, Kaihui and Gao, Peng and Jiang, Ying and Lin, Li and Zhao, Xiaoxu and Wang, Lifen and Bai, Xuedong and Fu, Wangyang and Wang, Jie-Yu and Li, Maozhi and Lei, Ting and Zhang, Yanfeng and Hou, Yanglong and Pei, Jian and Pennycook, Stephen J. and Wang, Enge and Chen, Ji and Zhou, Wu and Liu, Lei},
  journal   = {Nature},
  title     = {Disorder-tuned conductivity in amorphous monolayer carbon},
  year      = {2023},
  issn      = {1476-4687},
  month     = mar,
  number    = {7950},
  pages     = {56--61},
  volume    = {615},
  doi       = {10.1038/s41586-022-05617-w},
  publisher = {Springer Science and Business Media LLC},
}

@Article{Bai2024,
  author    = {Bai, Xiuhui and Hu, Pengfei and Li, Ang and Zhang, Youwei and Li, Aowen and Zhang, Guangjie and Xue, Yufeng and Jiang, Tianxing and Wang, Zezhou and Cui, Hanke and Kang, Jianxin and Zhao, Hewei and Gu, Lin and Zhou, Wu and Liu, Li-Min and Qiu, Xiaohui and Guo, Lin},
  journal   = {Nature},
  title     = {Nitrogen-doped amorphous monolayer carbon},
  year      = {2024},
  issn      = {1476-4687},
  month     = sep,
  number    = {8032},
  pages     = {80--84},
  volume    = {634},
  doi       = {10.1038/s41586-024-07958-0},
  publisher = {Springer Science and Business Media LLC},
}

@Article{Piatti2021,
  author    = {Erik Piatti and Adrees Arbab and Francesco Galanti and Tian Carey and Luca Anzi and Dahnan Spurling and Ahin Roy and Ainur Zhussupbekova and Kishan A. Patel and Jong M. Kim and Dario Daghero and Roman Sordan and Valeria Nicolosi and Renato S. Gonnelli and Felice Torrisi},
  journal   = {Nature Electronics},
  title     = {Charge transport mechanisms in inkjet-printed thin-film transistors based on two-dimensional materials},
  year      = {2021},
  month     = {dec},
  number    = {12},
  pages     = {893--905},
  volume    = {4},
  doi       = {10.1038/s41928-021-00684-9},
  publisher = {Springer Science and Business Media {LLC}},
}

@Article{Mondal2021,
  author    = {Mondal, Sandeep K. and Biswas, Ananya and Pradhan, Jyoti R. and Dasgupta, Subho},
  journal   = {Small Methods},
  title     = {Inkjet‐Printed MoS2 Transistors with Predominantly Intraflake Transport},
  year      = {2021},
  issn      = {2366-9608},
  month     = oct,
  number    = {12},
  volume    = {5},
  doi       = {10.1002/smtd.202100634},
  publisher = {Wiley},
}

@Article{Gabbett2024,
  author    = {Gabbett, Cian and Kelly, Adam G. and Coleman, Emmet and Doolan, Luke and Carey, Tian and Synnatschke, Kevin and Liu, Shixin and Dawson, Anthony and O’Suilleabhain, Domhnall and Munuera, Jose and Caffrey, Eoin and Boland, John B. and Sofer, Zdeněk and Ghosh, Goutam and Kinge, Sachin and Siebbeles, Laurens D. A. and Yadav, Neelam and Vij, Jagdish K. and Aslam, Muhammad Awais and Matkovic, Aleksandar and Coleman, Jonathan N.},
  journal   = {Nature Communications},
  title     = {Understanding how junction resistances impact the conduction mechanism in nano-networks},
  year      = {2024},
  issn      = {2041-1723},
  month     = may,
  number    = {1},
  volume    = {15},
  doi       = {10.1038/s41467-024-48614-5},
  publisher = {Springer Science and Business Media LLC},
}

@Article{Grillo2025,
  author    = {Grillo, Alessandro and Toral‐Lopez, Alejandro and Marian, Damiano and Pelella, Aniello and Peng, Zixing and Wang, Jingjing and Faella, Enver and Passacantando, Maurizio and Di Bartolomeo, Antonio and Fiori, Gianluca},
  journal   = {Advanced Functional Materials},
  title     = {Effects of Temperature Annealing on the Intrinsic Transport Mechanisms of Solution Processed Graphene Nanosheet Networks},
  year      = {2025},
  issn      = {1616-3028},
  month     = may,
  doi       = {10.1002/adfm.202501534},
  publisher = {Wiley},
}

@Article{Mott_1969,
  author    = {Mott, N. F.},
  journal   = {Philosophical Magazine},
  title     = {Conduction in non-crystalline materials: III. Localized states in a pseudogap and near extremities of conduction and valence bands},
  year      = {1969},
  issn      = {0031-8086},
  month     = apr,
  number    = {160},
  pages     = {835--852},
  volume    = {19},
  doi       = {10.1080/14786436908216338},
  publisher = {Informa UK Limited},
}

@Article{Sun2020,
  author    = {Jianbo Sun and Maurizio Passacantando and Maurizia Palummo and Michele Nardone and Kristen Kaasbjerg and Alessandro Grillo and Antonio Di Bartolomeo and Jos{\'{e}} M. Caridad and Luca Camilli},
  journal   = {Physical Review Applied},
  title     = {Impact of Impurities on the Electrical Conduction of Anisotropic Two-Dimensional Materials},
  year      = {2020},
  month     = {apr},
  number    = {4},
  pages     = {044063},
  volume    = {13},
  doi       = {10.1103/physrevapplied.13.044063},
  publisher = {American Physical Society ({APS})},
}

@Article{Rudra2021,
  author    = {Rudra, Moumin and Tripathi, H.S. and Dutta, Alo and Sinha, T.P.},
  journal   = {Materials Chemistry and Physics},
  title     = {Existence of nearest-neighbor and variable range hopping in Pr2ZnMnO6 oxygen-intercalated pseudocapacitor electrode},
  year      = {2021},
  issn      = {0254-0584},
  month     = jan,
  pages     = {123907},
  volume    = {258},
  doi       = {10.1016/j.matchemphys.2020.123907},
  publisher = {Elsevier BV},
}

@Article{Aghashahi2022,
  author    = {Aghashahi, Nooshin and Mohammadizadeh, Mohammad Reza and Kameli, Parviz},
  journal   = {Physica Scripta},
  title     = {Variable range hopping conduction mechanisms in reduced rutile TiO2},
  year      = {2022},
  issn      = {1402-4896},
  month     = mar,
  number    = {4},
  pages     = {045408},
  volume    = {97},
  doi       = {10.1088/1402-4896/ac576b},
  publisher = {IOP Publishing},
}

@Article{Lee2024,
  author    = {Lee, Jonghoon and Ferguson, John B. and Hubbard, Amber M. and Ren, Yixin and Nepal, Dhriti and Back, Tyson C. and Glavin, Nicholas R. and Roy, Ajit K.},
  journal   = {Materials Today Communications},
  title     = {Transitions of electrical conduction mechanism in graphene flake van der Waals thin film},
  year      = {2024},
  issn      = {2352-4928},
  month     = jun,
  pages     = {108859},
  volume    = {39},
  doi       = {10.1016/j.mtcomm.2024.108859},
  publisher = {Elsevier BV},
}

@Article{Ortuno2021,
  author    = {Miguel Ortu{\~{n}}o and Francisco Estell{\'{e}}s-Duart and Andr{\'{e}}s M. Somoza},
  journal   = {physica status solidi (b)},
  title     = {Numerical Simulations of Variable-Range Hopping},
  year      = {2021},
  month     = {nov},
  number    = {1},
  pages     = {2100340},
  volume    = {259},
  doi       = {10.1002/pssb.202100340},
  publisher = {Wiley},
}

@Article{Nenashev2013,
  author    = {Nenashev, A. V. and Jansson, F. and Oelerich, J. O. and Huemmer, D. and Dvurechenskii, A. V. and Gebhard, F. and Baranovskii, S. D.},
  journal   = {Physical Review B},
  title     = {Advanced percolation solution for hopping conductivity},
  year      = {2013},
  issn      = {1550-235X},
  month     = jun,
  number    = {23},
  pages     = {235204},
  volume    = {87},
  doi       = {10.1103/physrevb.87.235204},
  publisher = {American Physical Society (APS)},
}

@Article{Upreti2019,
  author    = {Upreti, Tanvi and Wang, Yuming and Zhang, Huotian and Scheunemann, Dorothea and Gao, Feng and Kemerink, Martijn},
  journal   = {Physical Review Applied},
  title     = {Experimentally Validated Hopping-Transport Model for Energetically Disordered Organic Semiconductors},
  year      = {2019},
  issn      = {2331-7019},
  month     = dec,
  number    = {6},
  pages     = {064039},
  volume    = {12},
  doi       = {10.1103/physrevapplied.12.064039},
  publisher = {American Physical Society (APS)},
}

@Article{Bhandari2023,
  author    = {Bhandari, Preeti and Malik, Vikas and Schechter, Moshe},
  journal   = {Physical Review B},
  title     = {Variable range hopping in a nonequilibrium steady state},
  year      = {2023},
  issn      = {2469-9969},
  month     = jul,
  number    = {2},
  pages     = {024203},
  volume    = {108},
  doi       = {10.1103/physrevb.108.024203},
  publisher = {American Physical Society (APS)},
}

@Article{Li2024,
  author    = {Li, Siyan and Yu, Jiayang and Zhu, Ye and Zhang, Tianyou},
  journal   = {Journal of Non-Crystalline Solids},
  title     = {Charge transport in disordered materials with exponentially distributed density of states},
  year      = {2024},
  issn      = {0022-3093},
  month     = sep,
  pages     = {123115},
  volume    = {640},
  doi       = {10.1016/j.jnoncrysol.2024.123115},
  publisher = {Elsevier BV},
}

@Article{Shlesinger2017,
  author    = {Shlesinger, Michael F.},
  journal   = {The European Physical Journal B},
  title     = {Origins and applications of the Montroll-Weiss continuous time random walk},
  year      = {2017},
  issn      = {1434-6036},
  month     = may,
  number    = {5},
  volume    = {90},
  doi       = {10.1140/epjb/e2017-80008-9},
  publisher = {Springer Science and Business Media LLC},
}

@Article{Wang2023,
  author    = {Wang, Jiejun and Zeng, Huizhong and Xie, Yiduo and Zhao, Zebin and Pan, Xinqiang and Luo, Wenbo and Shuai, Yao and Tang, Ling and Zhu, Dailei and Xie, Qin and Wan, Limin and Wu, Chuangui and Zhang, Wanli},
  journal   = {Advanced Intelligent Systems},
  title     = {Analog Ion‐Slicing LiNbO3 Memristor Based on Hopping Transport for Neuromorphic Computing},
  year      = {2023},
  issn      = {2640-4567},
  month     = jul,
  number    = {10},
  volume    = {5},
  doi       = {10.1002/aisy.202300155},
  publisher = {Wiley},
}

@Article{Tessler2009,
  author    = {Nir Tessler and Yevgeni Preezant and Noam Rappaport and Yohai Roichman},
  journal   = {Advanced Materials},
  title     = {Charge Transport in Disordered Organic Materials and Its Relevance to Thin-Film Devices: A Tutorial Review},
  year      = {2009},
  month     = {may},
  number    = {27},
  pages     = {2741--2761},
  volume    = {21},
  doi       = {10.1002/adma.200803541},
  publisher = {Wiley},
}

@Article{Coehoorn2012,
  author    = {Coehoorn, R. and Bobbert, P. A.},
  journal   = {physica status solidi (a)},
  title     = {Effects of Gaussian disorder on charge carrier transport and recombination in organic semiconductors},
  year      = {2012},
  issn      = {1862-6319},
  month     = dec,
  number    = {12},
  pages     = {2354--2377},
  volume    = {209},
  doi       = {10.1002/pssa.201228387},
  publisher = {Wiley},
}

@Article{Wu2017,
  author    = {Wu, Zhangting and Ni, Zhenhua},
  journal   = {Nanophotonics},
  title     = {Spectroscopic investigation of defects in two-dimensional materials},
  year      = {2017},
  issn      = {2192-8614},
  month     = mar,
  number    = {6},
  pages     = {1219--1237},
  volume    = {6},
  doi       = {10.1515/nanoph-2016-0151},
  publisher = {Walter de Gruyter GmbH},
}

@Article{Zheng2019,
  author    = {Zheng, Yu Jie and Chen, Yifeng and Huang, Yu Li and Gogoi, Pranjal Kumar and Li, Ming-Yang and Li, Lain-Jong and Trevisanutto, Paolo E. and Wang, Qixing and Pennycook, Stephen J. and Wee, Andrew T. S. and Quek, Su Ying},
  journal   = {ACS Nano},
  title     = {Point Defects and Localized Excitons in 2D WSe2},
  year      = {2019},
  issn      = {1936-086X},
  month     = may,
  number    = {5},
  pages     = {6050--6059},
  volume    = {13},
  doi       = {10.1021/acsnano.9b02316},
  publisher = {American Chemical Society (ACS)},
}

@Article{Sun2019,
  author    = {Sun, Tao and Zhang, Guoqiang and Xu, Dong and Lian, Xu and Li, Hexing and Chen, Wei and Su, Chenliang},
  journal   = {Materials Today Energy},
  title     = {Defect chemistry in 2D materials for electrocatalysis},
  year      = {2019},
  issn      = {2468-6069},
  month     = jun,
  pages     = {215--238},
  volume    = {12},
  doi       = {10.1016/j.mtener.2019.01.004},
  publisher = {Elsevier BV},
}

@Article{Hidaka2023,
  author    = {Hidaka, Atsuki and Kondo, Yuki and Takeshita, Akinobu and Matsuura, Hideharu and Eto, Kazuma and Ji, Shiyang and Kojima, Kazutoshi and Kato, Tomohisa and Yoshida, Sadafumi and Okumura, Hajime},
  journal   = {Japanese Journal of Applied Physics},
  title     = {Comparison of temperature-dependent resistivity of heavily Al- and N-codoped 4H-SiC grown by physical vapor transport and heavily Al-doped 4H-SiC grown by chemical vapor deposition},
  year      = {2023},
  issn      = {1347-4065},
  month     = oct,
  number    = {10},
  pages     = {101001},
  volume    = {62},
  doi       = {10.35848/1347-4065/acfb64},
  publisher = {IOP Publishing},
}

@Article{Matsuura2020,
  author    = {Matsuura, Hideharu and Takeshita, Akinobu and Hidaka, Atsuki and Ji, Shiyang and Eto, Kazuma and Mitani, Takeshi and Kojima, Kazutoshi and Kato, Tomohisa and Yoshida, Sadafumi and Okumura, Hajime},
  journal   = {Japanese Journal of Applied Physics},
  title     = {Sign of Hall coefficient in nearest-neighbor hopping conduction in heavily Al-doped p-type 4H-SiC},
  year      = {2020},
  issn      = {1347-4065},
  month     = may,
  number    = {5},
  pages     = {051004},
  volume    = {59},
  doi       = {10.35848/1347-4065/ab8701},
  publisher = {IOP Publishing},
}

@Article{Matsuura2019,
  author    = {Matsuura, Hideharu and Takeshita, Akinobu and Imamura, Tatsuya and Takano, Kota and Okuda, Kazuya and Hidaka, Atsuki and Ji, Shiyang and Eto, Kazuma and Kojima, Kazutoshi and Kato, Tomohisa and Yoshida, Sadafumi and Okumura, Hajime},
  journal   = {Japanese Journal of Applied Physics},
  title     = {Transition of conduction mechanism from band to variable-range hopping conduction due to Al doping in heavily Al-doped 4H-SiC epilayers},
  year      = {2019},
  issn      = {1347-4065},
  month     = sep,
  number    = {9},
  pages     = {098004},
  volume    = {58},
  doi       = {10.7567/1347-4065/ab3c2c},
  publisher = {IOP Publishing},
}
\printglossary[type=\acronymtype]

\end{document}